\documentclass[onecolumn,amsmath,amssymb,12pt,superscriptaddress,nofootinbib]{revtex4-1}
\pdfoutput=1

\usepackage[english]{babel}
\usepackage{amsthm}
\usepackage{amssymb}
\usepackage{amsmath}
\usepackage{amsthm}
\usepackage[]{graphicx}
\usepackage{tensor}
\usepackage{color}
\usepackage{framed}
\usepackage{framed}
\usepackage[bookmarks,linktocpage, colorlinks=true, plainpages = false, citecolor = blue,  linkcolor=blue, urlcolor = blue, filecolor = blue]{hyperref} 
\usepackage{natbib}
\usepackage{dcolumn}
\usepackage{enumerate}
\usepackage{bm}
\usepackage{graphicx}
\usepackage{caption}
\usepackage{subcaption}

\theoremstyle{definition}

\usepackage{tikz}
\usetikzlibrary{decorations.pathmorphing, patterns,shapes}

\input epsf

\begin{document}
\allowdisplaybreaks
\begin{titlepage}
\title{
Existence of real time quantum path integrals}
\author{Job Feldbrugge}
\email{job.feldbrugge@ed.ac.uk}
\affiliation{Higgs Centre for Theoretical Physics, James Clerk Maxwell Building, Edinburgh EH9 3FD, UK}
\affiliation{Department of Physics, Carnegie Mellon University, 5000 Forbes Ave, Pittsburgh, PA 15217, USA}
\author{Neil Turok}
\email{neil.turok@ed.ac.uk}
\affiliation{Higgs Centre for Theoretical Physics, James Clerk Maxwell Building, Edinburgh EH9 3FD, UK}
\affiliation{Perimeter Institute for Theoretical Physics, 31 Caroline St N, Waterloo, ON N2L2Y5, Canada}
\begin{abstract}
%\vspace{.25cm}
Many interesting physical theories have analytic classical actions. We show how Feynman's path integral may be defined non-perturbatively, for such theories, {\it without} a Wick rotation to imaginary time. We start by introducing a class of smooth regulators which render interference integrals absolutely convergent and thus unambiguous. The analyticity of the regulators allows us to use Cauchy's theorem to deform the integration domain onto a set of relevant, complex ``thimbles" (or generalized steepest descent contours) each associated with a classical saddle. The regulator can then be removed to obtain an exact, non-perturbative representation. We show why the usual method of gradient flow, used to identify relevant saddles and steepest descent ``thimbles" for  finite-dimensional oscillatory integrals, fails in the infinite-dimensional case. For the troublesome high frequency modes, we replace it with a method we call ``eigenflow" which we employ to identify the infinite-dimensional, complex ``eigenthimble" over which the real time path integral is absolutely convergent. We then bound the path integral over high frequency modes by the corresponding Wiener measure for a free particle. Using the dominated convergence theorem we infer that the interacting path integral defines a good measure. While the real time path integral is more intricate than its Euclidean counterpart, it is superior in several respects. It seems particularly well-suited to theories such as quantum gravity where the classical theory is well developed but the Euclidean path integral does not exist. 
\end{abstract}
\maketitle
\end{titlepage}
%\tableofcontents

%%%%%%%%%%%%%%%%%%%%%%%%%%%%%%%%%%%%%%%%%%%%%%%%%%%%%%%%%%%%%%%%%%%%%%%%%%%%%%%%%%%%
\section{Introduction}

Feynman's path integral formulation of quantum mechanical theories \cite{Feynman:1948,Feynman:1965} occupies a peculiar position in fundamental physics. On the one hand, it provides the most elegant known description of continuous quantum systems, most notably gauge theories including the standard model. It has been successfully applied in an innumerable variety of models and problems (Ref.~\cite{Grosche:1998} is a compendium for non-relativistic quantum mechanics). On the other hand, attempts to rigorously {\it define} the real time quantum path integral, as is required to directly describe dynamical phenomena, have failed for over half a century (for a review, see \cite{Klauder:2003,Klauder:2010}). The standard approach, dating back to Mark Kac in 1949~\cite{Kac:1949}, involves Wick rotating time to imaginary values and working in Euclidean spacetime. This approach has since dominated rigorous and non-perturbative approaches to path integrals and quantum field theory~\cite{Glimm:2012}. It has proven very fruitful, for example, in lattice gauge theory. However, it has the undesirable feature of replacing interference, which is fundamental to quantum physics, with statistical physics which has a qualitatively different character. The analytic continuation back to real time is often impractical. This problem has grown in urgency as experimental advances, across a vast range of fields, from biophysics to radio astronomy, have made dynamical quantum processes accessible including intricate interference behavior~\cite{Tannor:2007}. The Euclidean approach, furthermore, encounters problems even when describing equilibrium phenomena, in cases where the Euclidean action is not real and positive. Examples include dense baryonic matter~\cite{Alexandru:2020wrj} and quantum lattice models with a ``sign problem"~\cite{Mondaini:2021ywk} including the Hubbard model on a square lattice, thought to be a good model for exotic superconductivity. Our own work is motivated by the challenge of defining the path integral for gravity and cosmology~\cite{Teitelboim:1982, Teitelboim:1983, Wheeler:1987, Feldbrugge:2017, Feldbrugge:2017b, Tucci:2019, Turok:2022}.

Our goal in this paper is to take a step towards defining real time quantum path integrals, in the relatively simple setting of a non-relativistic particle in a potential. Instead of analytically continuing in time, we analytically continue in the dynamical variables. We generalize Picard-Lefschetz (or exact, steepest descent) analysis for finite-dimensional oscillatory integrals to the infinite-dimensional case. {\it We thereby obtain a fully non-perturbative expression for the real time path integral, in which regulators such as  Feynman's ``$i\epsilon$" are unnecessary}. Our arguments are not completely rigorous  but we feel they are sufficiently novel and convincing to be worth presenting in the hope that mathematicians can develop them into rigorous proofs. In a companion paper, we explicitly calculate the real time propagator in some illustrative examples, as a practical demonstration of the utility of these methods~\cite{FT2}. 

We introduce several novel concepts here. First, we define physically reasonable, smooth regulators which make oscillatory interference integrals such as real time path integrals mathematically meaningful. Provided the action is analytic in the dynamical variables, the analyticity of the regulator allows us to exploit Cauchy's theorem to deform the original, real integration domain onto a corresponding set of complex ``thimbles" (or generalized steepest descent contours) upon which the path integral is absolutely convergent. The regulator may then be completely removed. In this way, we map a highly oscillatory, regulated real time path integral onto a set of well-defined, regulator-independent infinite-dimensional measures (we provide an introduction to measure theory in Appendix~\ref{measuretheory}).  We will show that gradient flow, usually used in finite-dimensional Picard-Lefschetz analysis, fails in the infinite-dimensional case because it leads to ill-defined partial differential equations. Instead, we introduce a new method we call  ``eigenflow," which relies only on solutions to {\it ordinary} differential equations. Using eigenflow, we find the ``eigenthimble" for the high frequency modes, namely the half-dimensional subspace of the space of complexified paths, which plays the same role as a ``thimble"  in finite-dimensional Picard-Lefschetz analysis. We are able, in particular, to bound the high frequency part of the path integral for an interacting theory by the corresponding path integral for a free particle. Since the latter is associated with a well-defined, infinite-dimensional measure, closely related to the Wiener measure (see Appendix~\ref{measuretheory}), it follows from Lesbesgue's dominated convergence theorem  (see Appendix~\ref{theorems}) that the interacting, real time path integral also yields a well-defined measure. If this argument can be made rigorous, it would constitute a complete definition and proof of existence for real time path integrals in quantum mechanics. 

The real time propagator in non-relativistic quantum mechanics is given by:
 \begin{align}
K(x_1,t_1; x_0,t_0) \equiv  \langle x_1| \hat{U}(t_1,t_0) |x_0\rangle \Theta(t_1-t_0)  \sim \int_{x_0}^{x_1}  e^{i S[x]/\hbar} \,{\cal D} x,
\label{e00}
\end{align}
where  $\hat{U}(t_1,t_0)=T e^{-i\int_{t_0}^{t_1} \hat{H}(t) \mathrm{d}t/\hbar} $ is the time evolution operator, $\hat{H}(t)$ is the Hamiltonian operator and $|x_0\rangle$, $|x_1\rangle$  are position eigenstates. The propagator $K(x_1,t_1;x_0,t_0)$ may also be defined as the solution of the time-dependent Schr\"odinger equation with a delta function source, namely  $(i \hbar \partial_{t_1} -\hat{H}(x_1,t_1)) K(x_1,t_1;x_0,t_0)=i \hbar \delta(t_1-t_0)\delta(x_1-x_0)$, with the ``causal" boundary condition that $K(x_1,t_1;x_0,t_0)=0$, for $t_1<t_0$. Integrating the first equation over $t_1$, from just below $t_0$ to just above it ({\it i.e.}, from $t_0^-$ to $t_0^+$), the second implies that $\displaystyle \lim_{t_1\rightarrow t_0^+}K(x_1,t_1;x_0,t_0)=\delta(x_1-x_0)$, consistent with its definition in (\ref{e00}). The causal boundary condition ensures perturbative unitarity (see, {\it e.g.}, Refs.~\cite{Bjorken:100769,Feynman:1965}). 
(Note also that the operator appearing on the lhs of the defining equation is precisely the Hamiltonian constraint which generalizes neatly to the relativistic theory, and the rhs is already relativistically invariant.)

The last expression in (\ref{e00}) is, of course, Feynman's beautiful formula which has, however, remained only heuristically defined to this day. Our aim is to improve on this state of affairs. Feynman was well aware of the mathematical difficulties. In his textbook with Hibbs he states, of the discretized-time definition of the sum over paths, ``...we feel that the possible awkwardness of the special definition of the sum over all paths may eventually require new definitions to be formulated. Nevertheless, the concept of the sum over paths, like the concept of an ordinary integral, is independent of a special definition and valid in spite of the failure of such definitions" (Ref.~\cite{Feynman:1965}, p.34).  Having set aside rigor with these not entirely convincing arguments, they proceed to derive many very interesting results. In recent years, there has been a growing effort to develop an improved definition and understanding of real time path integrals (see, for example, \cite{Kent:2013,Donadi:2021yhd} and the reviews~\cite{Klauder:2003,Klauder:2010}).

In this paper, we shall start by explaining the mathematical subtleties of real time path integrals and then attempt to systematically resolve them. The outline is as follows. In Section \ref{smo} we explicitly exhibit the ambiguity in the simplest real time path integral, namely the discretized time path integral for a free, non-relativistic particle. We show how the ambiguity is resolved by introducing a class of physically reasonable, smooth regulators and then using contour deformation and appropriate limits to derive regulator-independent results. In Section \ref{form} we formally define path integrals in terms of an infinite number of eigenmodes. In Section \ref{eigenflow} we  show how gradient flow methods, usually used to identify generalized, higher dimensional steepest descent contours for interference integrals, fail in the infinite dimensional case. For the troublesome high frequency modes, we introduce a new method -- eigenflow -- to find the complex domain -- the eigenthimble -- over which the infinite-dimensional path integral is well-defined. In Section \ref{eigenthimble} we bound  the contour-deformed path integral over the high frequency modes by a well-defined measure related to the Wiener measure for a free particle. We provide a brief, informal summary of the relevant measure theory in Appendix~\ref{measuretheory} and key mathematical theorems in Appendix~\ref{theorems}. In Section \ref{quartic} we check these concepts explicitly for the quartic oscillator.  Appendix~\ref{etapp} provides analytic formulae for the steepest descent thimbles and the eigenthimbles in the lowest frequency approximation, for that example. Finally, Section \ref{advantages} summarizes some of the advantages of the real time, as opposed to the Euclidean, path integral. 

One of the surprises of our treatment is that it leads to an  {\it exact} representation of a path integral in which the classical theory plays an organising role. For example for a particle in a potential, with action $S[x]=\int_{t_0}^{t_1} \left( {1\over 2}M\dot{x}^2-V(x)\right) \mathrm{d}t$, the propagator may be written
 \begin{align}
K(x_1,t_1-t_0;x_0) =  \sqrt{M\over 2 \pi i \hbar (t_1-t_0)}\Theta(t_1-t_0) \sum\limits_{C}   e^{i S_C/\hbar}  n_C \int_{\mathcal{J}_{C}} e^{i \theta_{C}} d\mu_{C},
\label{e0}
\end{align}
(using time translation invariance to simplify the notation for $K$). The propagator implements unitary evolution via $\Psi(x_1,t_1) =\int K(x_1,t_1-t_0;x_0) \Psi(x_0,t_0) dx_0$. Hence,  it has dimensions of inverse length, which are accounted for by the square root prefactor in (\ref{e0}). 

The sum in (\ref{e0}) runs over {\it relevant classical} paths $x_{C}(t)$, real or complex, with action $S_C$, which satisfy the equations of motion and the boundary conditions. A precise criterion of relevance is provided by Picard-Lefschetz theory~\cite{Witten:2010, Basar:2013,Vassiliev}. For finite-dimensional integrals, every generic saddle $C$ ({\it i,e,} one where the hessian is non-degenerate) has a unique steepest {\it ascent} thimble $\mathcal{K}_{C}$ and a steepest descent thimble ${\cal J}_C$. With an appropriate choice of orientation, the intersection matrix of these thimbles is $\left({\cal K}_C, {\cal J}_{C'}\right)=\delta_{C,C'}$. If the original, oriented integration domain ${\cal C}$ can be deformed into a sum of descent thimbles $\sum\limits_{C}  n_C \mathcal{J}_{C}$, with $n_C$ an integer, it follows from the topological invariance of the intersection form that $n_C= \left({\cal K}_C, {\cal C}\right)$. Thus, $n_C$ is nonzero and a saddle is ``relevant" if and only if its steepest ascent thimble intersects the original contour. In our work, we will assume (and explicitly confirm in detailed calculations~\cite{FT2}) that this criterion still applies in the infinite-dimensional case. 

Note that, although we integrate over all fluctuations about every relevant classical saddle, there is no overcounting of paths because the complex thimbles $\mathcal{J}_{C}$ are entirely distinct. Upon every thimble the path integral over fluctuations about $x_{C}(t)$ converges absolutely and defines a positive, infinite-dimensional measure $d\mu_C$. The phase factor $e^{i \theta_{C}}$, with $\theta_C$ real, is a well-defined quantity at each point of the measure space, as we shall describe below. In the limit $\hbar \rightarrow 0$, the measure becomes tightly focused on the corresponding saddle and $e^{i \theta_{C}}$ tends to the well-known, semi-classical ``Maslow" phase factor~\cite{Keller, Maslow,Tannor:2007}, as we shall explain. We find it remarkable that there is an exact expression for the quantum mechanical propagator which, for any value of $\hbar$, is so clearly organized by the classical theory. In effect, by extremizing the interference amplitude over paths, the classical theory efficiently compresses the information contained in the quantum theory. 

With an eye to future applications in quantum gauge theories and, ultimately, gravity, we shall use the quartic oscillator as a concrete example. We explicitly construct the high frequency eigenthimble and show how the real time path integral efficiently describes non-perturbative phenomena which would be hard to recover from a Euclidean path integral. At the end of the paper, we provide an explicit example of the information compression provided by the classical theory, through the formula (\ref{e0}). In a companion paper, we implement a numerical method for determining the relevance (in the Picard-Lefschetz sense outlined above) of complex saddles, in the infinite-dimensional case. We also explore the detailed implementation of (\ref{e0}) in situations where both real and complex saddles are relevant~\cite{FT2}. 

Formulae closely analogous to (\ref{e0}) appear in fascinating recent discussions of ``resurgence,"  which emphasize the connections between the non-perturbative contributions to path integrals and perturbative contributions which may be much more easily derived. For a selection of recent papers, on a broad range of QFT and applied mathematical topics, see, {\it e.g.}, \cite{Dunne:2016nmc,Alexandru:2020wrj,Costin:2020, Bajnok:2021dri,Pazarbasi:2021ifb}. For an early study of perturbative saddle point methods, caustics and hyperasymptotics for both one- and multi-dimensional integrals, see \cite{Berry:1990, Berry:1991, Berry:1993, Howls:1997}. In our approach, all of the terms in the sum (\ref{e0}) formally originate, through analytic continuation, from a single regulated path integral. Whether this insight is ultimately useful for understanding the origin and meaning of ``resurgence" remains to be seen. 

\section{Smooth regulators and Cauchy's theorem}
\label{smo}

A real time path integral is, by definition, an interference integral: the quantum phase factor $e^{i S/\hbar}$ is integrated over the infinite, infinite-dimensional domain of real paths.  The integrand has modulus unity so the integral is not absolutely convergent. As is well-known in mathematics, such integrals are in general ambiguous and can depend, for example, on the coordinates chosen or the order in which partial integrals are performed (see Appendix~\ref{theorems} for examples). To avoid such ambiguities, one might try restricting the integral to a finite domain and taking the limit as the domain becomes large. Unfortunately, in more than one dimension this the result generally depends on the {\it shape} of the domain's boundary. To define path integrals, we must regulate extreme paths in a manner which allows them to interfere destructively and cancel out, which a sharp boundary does not. By instead regulating them smoothly, we implement the physical principle that the predictions for experiments on earth should not depend on paths which go via the moon. 

For integrals which are absolutely convergent, Fubini's theorem (see Appendix \ref{theorems}) shows that ambiguities of this type do not arise. We shall introduce a broad class of smooth regulators which are a) absolutely integrable and b) analytic so we can use Cauchy's theorem to transform the original, highly oscillatory integral into a sum of absolutely convergent integrals. After this transformation, the regulator can be removed. 

Consider the textbook example of the time-discretized path integral propagator for a free particle of mass $M$~\cite{Feynman:1965}. Take the amplitude for the particle to  ``go nowhere," {\it i.e.}, to travel from $x=0$ to $x=0$ in a positive time interval $T$. We can compare this with the solution to the time-dependent Schr\"odinger equation (or, equivalently, its Euclidean version -- the diffusion equation -- analytically continued back to real time) which gives the result $\sqrt{M/ (2 \pi i \hbar T)}$, where $\sqrt{i}=e^{i \pi/4}$. The discretized time path integral over ${\bf x}\equiv (x_1,\dots,x_N)$, the particle's position at each intermediate time, where $T=(N+1)\Delta t$ is, by definition
\begin{align}
\left({m\over 2 \pi i \hbar \Delta t}\right)^{N+1\over 2}\int \, e^{ i{M\over 2 \hbar \Delta t} \left(x_1^2+(x_2-x_1)^2\dots+(x_N-x_{N-1})^2+x_N^2\right)} \,\mathrm{d}^N {\bf x}. 
\label{e1.1}
\end{align}

Rescale ${\bm{x}}= {\bm{y}} \sqrt{\hbar \Delta t/M}$ to obtain a dimensionless integral with a single factor of $\sqrt{M/ (\hbar \Delta t)}$ out front. Write the real quadratic form appearing in the phase as ${1\over 2}{\bf y}^T {\bf K}{\bf y}$ and rotate ${\bm{y}}$ to diagonalize ${\bf K}$. Its diagonal elements $\lambda_n$, $n=1,\dots N$ are all positive, so rescale each $y_n$ by $\lambda_n^{-{1\over 2}}$. The product of eigenvalues $\det{\bf K}=\sqrt{N+1}$, from which we obtain the Schr\"odinger result multiplied by
\begin{align}
\int \prod_{l=1}^N {\, e^{i y_l^2} \mathrm{d}y_l \over \sqrt{\pi i }}.
\label{e1.2}
\end{align}
If each one-dimensional integral in this expression is defined with cutoffs, it indeed converges to unity as the cutoffs are taken to $\pm \infty$. This amounts to using a {\it box-shaped} cutoff for the $N$-dimensional integral and removing the cutoff by sending the box ends to $\pm \infty$ in all Cartesian directions. However, if we instead use a {\it spherical} cutoff at $r\equiv \sqrt{{\bf y}^2}=R$, after performing the angular integrals we obtain a completely different result:
\begin{align}
{ 2 \pi^{N\over 2}\over (i \pi)^{N\over 2} \Gamma(N/2)} \int_0^R  e^{i r^2} r^{N-1} \mathrm{d}r=1-{\Gamma(N/2,-i R^2)\over \Gamma(N/2)}.
\label{e1.3}
\end{align}
The last term is $\approx e^{i R^2} R^{N-2}/\left((i^{(N+2)/2} \Gamma(N/2)\right)$ at large $R$. For $N=2$ the result oscillates with fixed magnitude as $R\rightarrow \infty$ whereas for $N>2$ it grows in magnitude without bound. Thus, for all $N>1$, a sharp spherical cutoff fails to give a good limit.

If, however, we regulate (\ref{e1.2}) smoothly by inserting a factor $e^{-r^2/L^2}$ under the integral, it gives $ (1+i/L^2)^{-{N/2}}$ which has the correct limit as $L\rightarrow \infty$. The smooth cutoff gives a sensible physical result precisely because it allows extreme paths to interfere destructively and cancel out. To define an infinite class of smooth regulators, and to see why they all give the same result as the regulator is removed, we now turn to complex analysis.

What general properties should a smooth regulator $f(\bm{x})$ have? Clearly, it must be absolutely integrable, {\it i.e.}, $\int \mathrm{d}{\bm{x}} | f({\bm{x}})|$ must be finite,  so that its insertion renders the oscillatory integral over a pure phase absolutely convergent. This implies a bound on the Fourier transform, $\tilde{f}({\bm{k}}) =\int \, e^{i{\bm{k}}\cdot{\bm{x}}} f({\bm{x}})\mathrm{d}{\bm{x}}$ because $|\tilde{f}({\bm{k}})|\leq \int | f({\bm{x}})|\mathrm{d}{\bm{x}}$, for all ${\bm{k}}$. Second, a smooth regulator must tend to unity, $f({\bm{x}})\rightarrow 1$ at each ${\bm{x}}$, in some limit, in which it can be said to be removed. Finally, to allow us to use Cauchy's theorem, $f(\bm{x})$ should be analytic in the region of the complex ${\bm{x}}$-domain through which we need to deform the ${\bm{x}}$-integration contour. As an example, consider functions $f({\bm{x}})$, with $f({\bm{0}})=1$, whose Fourier transform $\tilde{f}({\bm{k}})$ is supported on a compact domain $D$ including the origin. Suppose, further, that $\tilde{f}({\bm{k}})$ is $C^\infty$ and that all its ${\bm{k}}$-derivatives vanish on $\partial D$.  Finally, assume $D$ lies entirely within an open ball,  $|{\bm{k}}|<L^{-1}$. Then, as $L\rightarrow \infty$, $\tilde{f}({\bm{k}})$ tends to a delta function and $f({\bm{x}})$ tends to unity at all finite ${\bm{x}}$.  By repeated integrations by parts, one can show that $|f({\bm{x}})|$ is bounded by a constant times any negative power of ${\bm{x}}^2$. Combined with the boundedness of $f({\bm{x})}$, which follows from that of $f({\bm{k}})$, it follows that $f({\bm{x}})$ is absolutely integrable. As we shall see shortly, a Paley-Wiener theorem implies that $f({\bm{x}})$ is also entire and bounds its growth off the real ${\bm{x}}$-domain by an exponential. For the oscillatory integrals of interest, this bound will suffice to ensure a good $L\rightarrow\infty$ limit in which the regulator is removed.

For example, consider the integral (\ref{e1.2}), with a smooth regulator of the above form inserted. Cauchy's theorem allows us to deform each $x_l$ integration contour from the real $x_l$-axis to (i) a portion of the steepest descent contour from the saddle at $x_l=0$: $x_l = e^{i\pi/4}w_l$, with $-R <w_l <R$, (ii) two arcs of radius $R$, on either side of the origin, returning the contour to the real $x_l$-axis, and (iii) the real line segments $(-\infty,-R]$ and $[R,\infty)$. A simple Paley-Wiener theorem (see~\cite{Strichartz}, Theorem 7.2.1) yields  $\displaystyle | f({\bm{x}})| <  e^{|{\rm Im} ({\bm{x}})|/L} \int_{D}\, |\tilde{f}({\bm{k}})|\mathrm{d}{\bm{k}}$. Along (i), our integrand falls off as a Gaussian, so this contribution converges absolutely in the limit $R\rightarrow \infty$. The integrals in (iii) tend to zero in the same limit because $f({\bm{x}})$ falls off faster than any negative power of ${\bm{x}}^2$. Finally, using $\displaystyle |{\rm Im} ({\bm{x}})|<\sum_{l=1}^N |{\rm Im} (x_l)| $ and setting $x_l\equiv Re ^{i \theta}$ along either arc in (ii), the magnitude of each $x_l$-integral is bounded by $R\int_0^{\pi/4}  e^{-R^2 \sin 2 \theta+(R/L) \sin \theta}\mathrm{d} \theta$. For $R\gg L$, the exponent is bounded above by $R^2$ times a negative linear function of $\theta$. Integrating the bounding function over $\theta$ bounds the integral by a constant times $R^{-1}$.    Thus, at fixed $L$, we can take the limit $R\rightarrow \infty$ and all that remains is the integral over the thimble defined by $ x_l =e^{i\pi/4}w_l$, with $w_l$ real.   Since the integral is now absolutely convergent, the dominated convergence theorem (see Appendix \ref{theorems}) allows us to take the limit $L\rightarrow \infty$ inside the integral. {\it Notice that the order of limits is crucial:} we hold $L$ fixed as we deform the contour and send $R\rightarrow \infty$. Only after taking that limit do we send $L\rightarrow \infty$ and remove the regulator. We have explicitly illustrated the procedure for a quadratic exponent, but the argument is easily generalized to higher powers.

To summarize: the notion of a smooth regulator allows us to map a large class of oscillatory integrals onto absolutely convergent integrals which require no regulator at all. For this to be possible, both the integrand and the regulator must be analytic in the complex domain through which the integration contour is deformed. Second, the integrand must decay rapidly enough at infinity, in that domain, that no contributions arise ``from infinity." When both conditions are met, extreme paths cancel out and a unique result is obtained. A suitable class of smooth regulators can likewise be defined in the infinite-dimensional case, but we shall defer a detailed discussion to future work. 

We have shown that making sense of the time-discretized path integral is a delicate matter, even for a free particle. When interactions are included, the task becomes exponentially more difficult. For example, for a polynomial potential $V(x)$ which has highest power $x^{2 P}$, with $P$ a positive integer, the time-discretized action with $N$ intermediate times is a polynomial of order $2P$ in $N$ variables. It follows that the discretized action has an exponentially large number -- precisely,  $(2P-1)^N$ -- saddles (real or complex). Discretizing the time imposes an upper limit on the energies of states one can describe: the number of states grows at most as a power law in $N$. Therefore, it is clear that the vast majority of saddles in the discretized theory can have nothing do with real physics. To be sure, for every saddle (real or complex) in the continuum theory, there should be a sequence of saddles in the discretized theory, which converges to the continuum saddle as $N$ is taken to infinity. But only saddles which are part of such sequences have anything to do with continuum physics. Since we want to use saddle point and steepest descent methods, we will therefore avoid using the time-discretized definition of the path integral. Instead, we shall first identify the relevant saddles (either real or complex) in the continuum theory, then give a continuum definition of the complex thimble associated with each saddle and only finally, when we need to actually calculate the path integral over the thimble, resort to a discrete approximation (such as restricting the space of paths to discrete times or a finite number of Fourier modes). The latter approximations are then reasonable because, at this final stage, we are dealing with an absolutely convergent integral.

\section{formal path integral and contour deformation} 
\label{form}

Consider a particle in a potential $V(x)$, with action $S[x]=\int_0^1 \left( {1\over 2} M T^{-1} \dot{x}^2- T V(x)\right) \mathrm{d}t$. We have rescaled the time $t$ so that $0\leq t\le1$, with $T\equiv t_1-t_0>0$. We want to integrate over all paths $x(t)$ satisfying $x(0)=x_0$, $x(1)=x_1$. Let $x(t)=x_C(t)+\delta x(t)$, with $x_C(t)$ a {\it relevant} classical solution (we discussed the criterion of ``relevance" below Eqn. (\ref{e0})).  We write the action as $S[x_C(t)+\delta x(t)]\equiv S_C+\delta S$, with $\delta S= \delta S^{(2)}[\delta x(t)]+\delta S^{({ho})}[\delta x(t)]$ including the quadratic and higher order terms,

\begin{align}
\delta S^{(2)}[\delta x(t)]&= \int_0^1{1\over 2} \left(\delta x\hat{O} \delta x\right)\, \mathrm{d}t, \cr
 \delta S^{({ho})}[\delta x(t)]  &= \int_0^1\left(-\sum_{k=3}^\infty {T\over k!}\delta x^k{\partial^k V\over \partial x^k}(x_C(t)) \right) \mathrm{d}t ,
\label{g1}
\end{align}
where $ \hat{O} \equiv -M T^{-1} {\mathrm{d}^2\over \mathrm{d}t^2 }- T {\partial^2 V\over \partial x^2}(x_C(t))$ plays the same role as the hessian for saddles in the finite-dimensional case. On occasion we shall refer to $ \hat{O}$ as the hessian operator. It also defines the Jacobi equation, $\hat{O}\delta x(t)=0$, whose solutions determine the functional determinant which the path integral over fluctuations yields at leading semiclassical order (we give simple examples in Section \ref{advantages}). 

Of course, $\delta S$ contains no term linear in the fluctuation $\delta x(t)$ because, by definition, $x_C(t)$ is a classical solution. Furthermore, since $x_C(t)$ satisfies the boundary conditions, $\delta x(t)$  must vanish at the initial and final times: $\delta x(0)=\delta x(1)=0$. The operator $ \hat{O}$ is symmetric: for real classical solutions, it is Hermitian. For complex classical solutions, one can find a basis in which $ \hat{O}=\hat{A}+i \hat{B}$ where $\hat{A}$ and $\hat{B}$ are real, symmetric operators which commute and can hence be simultaneously diagonalized (for complex, symmetric matrices this result is known as Autonne-Takagi factorization). In order to streamline our discussion, we shall for the most part ignore this complication, writing $\hat{O}$ for the operator which we diagonalize. However, nothing in our procedure restricts us to real classical solutions. 

Near the saddle, the quadratic action $\delta S^{(2)}$ dominates over $\delta S^{(ho)}$. It is therefore natural to express $\delta x(t)$ as  $\displaystyle \sum_{m=1}^\infty \delta x_m \psi_m(t)$, where the $\psi_m(t)$ are orthonormal eigenmodes of $\hat{O}$ (or $\hat{A}$ for complex solutions), ordered by their eigenvalues $\lambda_m$, for $m=1,2\dots \infty$. Standard theorems (see, {\it e.g.}, Ref.~\cite{MorseandFeshbach}) assure us that, as the mode number $m\rightarrow \infty$,  both $\psi_m(t)$ and $\lambda_m$ tend to the eigenmodes and eigenvalues of the corresponding free particle operator $ \hat{O}_{0} \equiv -(M/T) {\mathrm{d}^2\over \mathrm{d}t^2 }$, {\it i.e.}, $\psi_{0,m}(t)=\sqrt{2} \sin \pi m t$ and $\lambda_{0,m}=(\pi m)^2 (M/T)$.

Given a relevant classical solution $x_C(t)$, the corresponding contribution to the path integral propagator may be defined formally as an integral over all mode amplitudes:
\begin{align}
K_C(x_1,T;x_0)\equiv { \displaystyle  \sqrt{M\over 2 \pi i \hbar T}  e^{i {M (x_1-x_0)^2\over 2 \hbar T}} 
 \over \displaystyle \int  e^{i{S_0[x(t)]\over \hbar}}  \prod_{m=1}^\infty\mathrm{d}\delta x_m} \int  e^{i{S[x(t)]\over \hbar}  } \prod_{m=1}^\infty\mathrm{d}\delta x_m,  
 \label{g1a}
\end{align}
 where we again for simplicity assumed time translation invariance. The first factor may be thought of as a normalization constant. It is completely determined by the free particle path integral which, being Gaussian, is easily defined and performed. The free particle hessian operator $\hat{O}_0$, appearing in the free particle action $S_0[x(t)]$, has orthonormal eigenfunctions $\psi_{0,m}(t)=\sqrt{2} \sin (m \pi t)$ and eigenvalues $\lambda_{0,m}= {1\over 2} (M/T)(m \pi)^2 $. Any path $x(t)$ can be expressed as the sum of a classical solution and a fluctuation $\displaystyle \delta x(t)= \sum\limits_{m=1}^\infty  \delta x_m \psi_{0,m}(t)$. In the second factor, we do the same but for the interacting action and the operator $\hat{O}$, with orthonormal eigenfunctions $\psi_{m}(t)$ and eigenvalues $\lambda_{m}$. We write $\displaystyle \delta x(t)= \sum\limits_{m=1}^\infty  \delta x_m \psi_m(t)$ so that $\delta x_m=\int_0^1 \psi_m(t) \delta x(t) \mathrm{d}t$.  Formally, regarding $\int_0^1 \mathrm{d}t$ as a sum over a ``matrix index" $t$,  $\hat{O}_t\delta(t-t')$ is a symmetric ``matrix" and the integral operator $\int_0^1 \psi_m(t)  \mathrm{d}t $ is an ``orthogonal rotation" from $\delta x(t)$  to the $\{ \delta x_m, m=1,\dots \infty \}$, as the relation $\int_0^1  \psi_m(t) \psi_n(t) \mathrm{d}t =\delta_{m n}$ implies. The Jacobian of such a rotation is unity. Thus the infinite products appearing in the numerator and the denominator are formally the same whether $\delta x(t)$ is expressed in terms of the eigenmodes of $\hat{O}_0$ or $\hat{O}$. The same normalization constant applies in the free or the interacting propagator because both have the same short time limit, $\displaystyle \lim_{T\rightarrow 0^+} K(x_1,T;x_0) = \delta (x_1-x_0)$. 
 
To perform the free particle path integral, we first find the classical solution and its action,  $S_{0,C}={1\over 2} (M/T) (x_1-x_0)^2$. This cancels with the corresponding term in the numerator. The fluctuation action is $\displaystyle \delta S_0={1\over 2} \sum_{m=1}^\infty \lambda_{0,m} \delta x_m^2$. To integrate over the $\delta x_m$, we rotate the contour by setting $\delta x_m= e^{i {\pi\over 4}} w_m$  with $w_m$ real.  The space of real $w_m$ is the free particle thimble ${\cal J}_0$: we call this the Wiener thimble and the $w_m$ Wiener coordinates. After rotating to the Wiener thimble, we evaluate the free particle path integral, ignoring contributions ``from infinity" as justified in the previous section. The free particle path integral is then proportional to 
\begin{align}
\prod_{m=1}^\infty e^{-{w_m^2\over 2\sigma_m^2}} {\mathrm{d} w_m\over \sqrt{2 \pi} \sigma_m}\equiv  \mathrm{d}\mu_B ; \qquad \sigma_m= {\sqrt{\hbar T/M}\over m \pi} \equiv {W \over m \pi}.
 \label{g1w}
\end{align}
which is the ``Brownian bridge" probability measure (with weight $W$). This is a version of the Wiener measure where the paths are constrained to end at zero (for further details, see Appendix \ref{measuretheory}). Substituting (\ref{g1w}) into (\ref{g1a}), the interacting path integral becomes
\begin{align}
K_C(x_1,T;x_0)= \sqrt{M\over 2 \pi i \hbar T}e^{i{S_C\over \hbar}}  \int  e^{i{\delta S\over \hbar}} \prod_{m=1}^\infty {\mathrm{d} \delta x_m\over \sqrt{2 \pi i } \sigma_m },  
 \label{g1a2}
\end{align}

We emphasize that (\ref{g1a2}) is {\it not} yet mathematically meaningful. It will only become so when we have defined the complex domain or ``thimble"  ${\mathcal J}_C$ over which the fluctuations $\delta x_m$ are to be integrated. Even then, care must be taken {\it not} to interpret the product $\displaystyle \prod_{m=1}^\infty {\mathrm{d} \delta x_m/ (\sqrt{2 \pi i } \sigma_m )}$ as in any meaningful sense a ``measure:" it is not, as we explain in Appendix \ref{measuretheory}. Only when this factor is combined with the factor $e^{i{\delta S\over \hbar}}$ and evaluated on ${\cal J}_C$, do we obtain a good measure. By normalizing the interacting path integral as we have done, relative to the free path integral, functional determinants are also naturally defined in terms of ratios of infinite products of free and interacting eigenvalues. We shall discuss simple examples of such functional determinants in section \ref{advantages} below. In a companion paper, we implement the more elegant mathematical treatment of Refs.~\cite{Levit:1977, Forman:1987,McKane:1995} which rests upon the same free particle normalization. 

To deal with the remaining path integral in (\ref{g1a2}) we again exploit a simplification which occurs at short times or high frequencies. In this limit, we expect the kinetic term to dominate over the potential and the interacting ``thimble" to be close to that for the free particle. With this simplification in mind, we separate the fluctuation into low and high frequency parts, $\delta x(t)  =\delta x^{L}(t)+\delta x^{H} (t)$, where $\delta x^{L}$ and $\delta x^{H}$ are linear combinations, respectively, of modes $\psi_m(t)$ with $m< m_C$ and  $m\geq m_C$, where $m_C$ demarcates the boundary between low frequency, ``light" modes, including those involved in the classical solution, and high frequency, ``heavy" modes, which are almost completely unexcited in the classical solution. Write the fluctuation action as $\delta S=\delta S^L+\delta S^H$, with $\delta S^L$ comprising the terms not involving the high frequency modes and $\delta S^H$ the remainder. Then
\begin{align}
K_C(x_1,T; x_0)=    {e^{i{S_C\over \hbar}} \over \sqrt{2 \pi i \hbar T}}  \int   e^{i{\delta S^L\over \hbar }}    \prod_{m=1}^{m_C-1} {\mathrm{d} \delta x_m\over \sqrt{2 \pi i} \sigma_m}  
\left(\int  e^{i{\delta S^H\over \hbar}} \prod_{m=m_C}^{\infty} {\mathrm{d} \delta x_m\over \sqrt{2 \pi i} \sigma_m } \right).
 \label{g1b}
\end{align}
Again, we expect the infinite-dimensional measure, defined by the last, bracketed factor, to only make sense when both the phase factor and the infinite product are combined, as they are in the Brownian bridge. 

In the next section, we deform the contour of integration in the last factor onto what we call the ``high frequency eigenthimble" ${\cal J}^H$, upon which the high frequency path integral converges absolutely. Near the saddle, the eigenthimble is determined by the quadratic operator $\hat{O}$. As the frequency $m$ is raised, the kinetic term dominates over the potential and  the high frequency part of the eigenthimble ${\cal J}^H$ closely follows that for the free particle thimble ${\cal J}_0^H$, further and further out in the complex $\delta x_m$-plane. Writing $\delta x_m=e^{i \pi/4} w_m$, the contribution of the $m$'th mode to $i\delta S/\hbar$ is $\approx -w_m^2/(2 \sigma_m^2)$, out to greater and greater $w_m$, where the integrand is exponentially suppressed. The contribution of each additional high frequency mode to the last bracket in (\ref{g1b}) thus tends to unity so the product of these contributions has a finite limit as more and more modes are included.  In the following sections, we shall bound the path integral over ${\cal J}^H$ by the high frequency part of the Brownian Bridge, $d\mu_{B}^{H}$, thereby showing that it yields a well-defined, infinite-dimensional measure. Having defined the path integral measure for high frequency modes, the remaining finite-dimensional path integral over low frequency modes can be defined and performed using finite-dimensional Picard-Lefschetz methods (see, for example, \cite{Feldbrugge:2017b, Feldbrugge:2019arXiv190904632F, 2020arXiv201003089F}).

\section{Eigenflow and the eigenthimble} 
\label{eigenflow}

For finite-dimensional oscillatory integrals, the steepest descent or ascent contours from a saddle are usually found by gradient flow, following the set of falling or rising directions defined by the hessian at the saddle. As we shall see momentarily, gradient flow fails badly in the infinite-dimensional case. Our focus here will be on showing that the integration domain for the path integral over high frequency modes may be deformed onto a complex thimble upon which the path integral defines a good measure. Happily, in this simpler context, we are able to identify and implement an alternative to gradient flow. Our nonlinear method generalizes the analysis of the eigenfunctions of the hessian operator $\hat{O}$ which, as we have discussed, simplifies for high frequencies and naturally defines the descending and ascending directions in the complexified space of paths in the vicinity of a classical saddle. Hence, we call our method ``eigenflow." We will use eigenflow to find the ``eigenthimble" which plays the same role as the thimble does in the finite-dimensional setting, namely that of defining a half-dimensional space within the space of complexified integration variables, upon which the integral becomes absolutely convergent. To illustrate the method, it will suffice to discuss the amplitude to ``go nowhere" in a theory with a trivial classical saddle $x_C(t)=0$, whose classical action $S_C$ is zero.  Since, in this case, $\delta x(t)=x(t)$ and $\delta S[\delta x]=S[x]$, we can drop the $\delta$'s and thereby minimize notational clutter. From this point in the paper onwards, all quantities are fluctuations unless explicitly stated otherwise. 

The real part of the exponent in the path integral (\ref{g1a2}), which governs the magnitude of the integrand, defines the ``height function" $h[x]\equiv {\rm Re} \left[i S[x]\right]$. The kinetic term is diagonalized by $x(t)= e^{i \pi/4} (F(t)+i G(t))$, with $F(t)$ and $G(t)$ real:  $ {\rm Re} \left[ i\int_0^2  \,\dot{ x}^2\mathrm{d}t\right]=\int_0^1  {1\over 2} (-\dot{F}^2+\dot{G}^2)\mathrm{d}t$.
Clearly, as far this term is concerned, $F$ defines the steepest descent directions and $G$ the  steepest ascent directions. To proceed using gradient flow, we functionally differentiate $h$ with respect to $F$ and $G$. With the inclusion of potential terms, the gradient flow descent equations are pdes:
\begin{align}
\partial_\tau F&=-{\delta h\over \delta F}=-\partial_t^2{F}+{\rm nonlinear\, terms}, \label{g1c}\\
\partial_\tau G&=-{\delta h\over \delta G}=+\partial_t^2{G}+{\rm nonlinear \, terms}.\label{g1c2}
\end{align}
The second equation is a nonlinear diffusion equation, with $\tau$ playing the role of ``time" and $t$ ``space": the double derivative term damps away curvature, {\it i.e.}, the double derivative in $t$ in the function $G(t)$. However, the first equation has the opposite character. The double derivative term now {\it amplifies} curvature in $F(t)$ an uncontrolled manner: the higher the frequency, the faster $F$ grows.  Given a nontrivial path $F(t)$ and $G(t)$ at $\tau=0$, the nonlinear terms generate higher and higher frequency modes which grow faster and faster in $\tau$. There is a cascade to infinite frequency in arbitrarily short time, so the equations (\ref{g1c}, \ref{g1c2}) fail to have a well-defined solution.

The problem with gradient flow is that it is, in a sense, {\it too} effective. In the infinite-dimensional case, there are infinitely steep directions and these make the notion of a ``flow" ill-defined. What we need is a more selective flow, which operates ``mode by mode,"  and ``filters out" higher frequency modes. We shall define such  a flow momentarily but, before doing so, let us say a little more about the context in which we intend to apply it. For simplicity, consider quantum mechanical models where the potential $V(x)$ is a polynomial in $x$ which is bounded below so that the classical solutions are all regular. Clearly, at large complex $|x|$, the highest power dominates. In the remainder of the paper, we shall focus on proving that the real time path integral for such theories exists. 

For such potentials, at high frequencies and near the saddle, the kinetic term in the action dominates over the potential term. As $m$ is raised, the kinetic term dominates further and further out into the complex $x_m$-plane. Only when  $|x_m|$ becomes very large (for example, $|x_m|\sim m$ in the case of a quartic potential) do the potential terms start to compete. When they do, it is the highest power in $V(x)$ which dominates, so that all sub-leading powers may be neglected. We therefore reach the important conclusion that the high-frequency ``eigenthimble" we seek is determined by a balance between the kinetic term and the {\it highest} power in the potential.  This is an enormous simplification: not only can we neglect all sub-leading powers, we can neglect any dependence on the specific classical saddle and hence on the initial and final condition. We can focus on the amplitude to ``go nowhere" and the problem of determining the eigenthimble becomes {\it scale-invariant}. This allows us to calculate the eigenthimble for all (sufficiently high) frequencies, at once. We shall do so explicitly, in section \ref{quartic}, for the quartic oscillator. 

\begin{figure}
\centering
\includegraphics[width = 0.8 \textwidth]{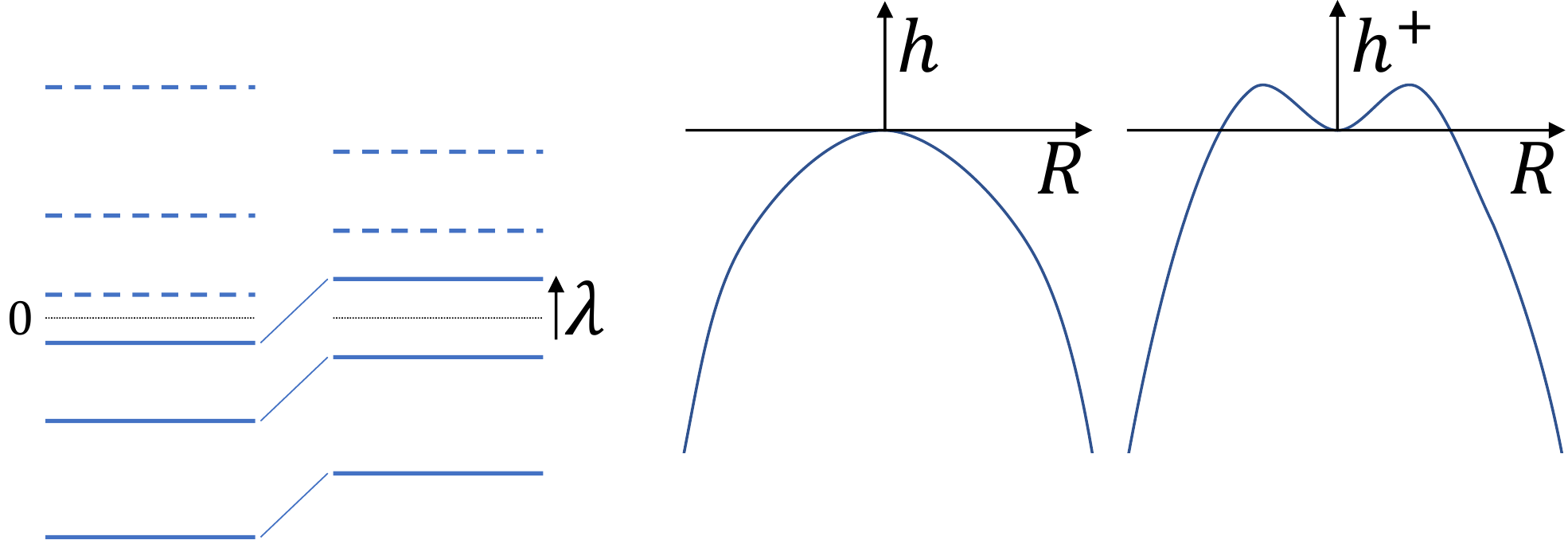}
\caption{The eigenvalues of the operator $\hat{O}$ governing fluctuations about a saddle come in equal and opposite pairs: adding  the Lagrange multiplier $\lambda$ raises all eigenvalues equally (left). Provided the potential falls faster than quadratically, at large $R^2$, when $\lambda$ sends an eigenvalue positive, two new saddles form. As $\lambda$ is increased, they move further out (right). }
\label{fig:diags}
\end{figure}

Let us now describe the eigenflow method. Since the height function $h$ is the real part of an analytic function, it follows from the Cauchy-Riemann equations that the eigenvalues of the operator $\hat{O}$ governing quadratic fluctuations, come in equal and opposite pairs (see Fig.~\ref{fig:diags}). Starting from a saddle, the descending directions are spanned by the eigenfunctions of $\hat{O}$ with positive eigenvalues $\lambda_m$.  The ``shallowest" of these directions is associated with the smallest such $\lambda_m$. To define the eigenflow downwards in the corresponding direction, we impose a constraint on the ``radius squared" $R^2$ of the fluctuation from the saddle using a positive Lagrange multiplier $\lambda$, and seed the flow with a solution which tends to a multiple of the corresponding eigenfunction of $\hat{O}$ as $\lambda$ approaches $\lambda_m$ from above. We thus replace the height function $h$ with $h^+[x]\equiv  h[x] +{1\over 2}\lambda( \int_0^1  |x(t)|^2 -R^2)\mathrm{d}t$. Notice that the additional term, being rotational symmetric in the complex $x$-plane, does not prejudice the direction of downward descent. Near the saddle, this term lifts all the eigenvalues of $-\hat{O}$, the operator appearing in the exponent after rotation to Wiener coordinates, by exactly the same amount. As mentioned above, to describe the high frequency eigenthimble, all we need to consider is the leading power of $x$ in the potential $V(x)$, at large $x$.  In the complex $x$-plane, the potential contributes to the height function as $\int_0^1 {\rm Re}[-i V(x) ]\mathrm{d}t$. This contribution runs to large negative values as the path $|x(t) |\rightarrow \infty$ in certain domains of the complex $x-$plane. As the Lagrange multiplier $\lambda$ is increased, the smallest negative eigenvalue of $-\hat{O}$ becomes positive and two new saddles appear, at small $R^2$ (see Fig.~\ref{fig:diags}). These saddles are extrema of $h^+[x]$: treating $x(t)$ and its complex conjugate $\overline{x}(t)$ as independent variables and taking functional derivatives (\`a la Wirtinger), we obtain the eigenflow equations: 
\begin{align}
&{\delta h\over  \delta x}=-{1\over 2} \lambda \overline{x}, \qquad {\delta h\over  \delta \overline{x}}=-{1\over 2} \lambda x,
\label{g1ca}
\end{align}
with $\lambda>\lambda_m$ for the $m$'th descent eigenflow from a saddle. 
In the corresponding saddle point solutions, at large $\lambda$, $x$ spends most of the time $0<t<1$ near the potential maxima at large $R^2$.  As $\lambda$ is raised further, the saddles move out to larger $R^2$ (see Fig.~\ref{fig:diags}).  When the next negative eigenvalue crosses zero, two more saddles appear, in an orthogonal direction in function space, and move outwards, and so on.  

 In finite-dimensional gradient flow as applied, for example, to an interference integral over a set of variables ${\bm{x}}$, one separates the exponent into its real and imaginary parts via $iS({\bm{x}})= h({\bm{x}})+i H({\bm{x}})$. The descent thimble may be found by solving the flow equation $\partial_\tau {\bm{x}}=-(\delta h/  \delta \overline{\bm{x}})$ for the complexified coordinate ${\bm{x}}$. As a result of the analyticity of $S({\bm{x}})$, it follows that $H({\bm{x}})$ is constant along such flows. The same is {\it not} true of eigenflow as defined in (\ref{g1ca}). The fact that $H$ is not constant does not affect the absolute convergence of the integral, since $H$ does not alter the magnitude of the integrand. However, the fact that $H$ varies means that we must keep track of it when computing the integral over the eigenthimble. The phase $e^{i H/\hbar}$ comprises one contribution to the phase $e^{i \theta_C}$ in the formula (\ref{e0}): the other arises from the Jacobian of the change of variables from $x_m$ to the natural coordinates on the eigenthimble, as we shall explain.

For every positive integer $m\geq m_C$ and for $\lambda$ just above $\lambda_m$, at small $R^2=\int_0^1  |x(t)|^2\mathrm{d}t$ there are two solutions $\delta x \approx e^{i \pi/4}R \psi_m(t)$, with $R$ either positive or negative.  As $\lambda$ increases, it remains in one to one correspondence with $R^2$ and the height function $h$ becomes more and more negative.  Using (\ref{g1ca}), we can easily obtain an upper bound on $h$:
\begin{align}
&{\mathrm{d} h\over \mathrm{d}\lambda} = \int_0^1 \, \left({\mathrm{d}  x \over \mathrm{d} \lambda} {\delta h \over \delta x}+{\mathrm{d} \overline{x} \over \mathrm{d} \lambda} {\delta h \over \delta \overline{x}}\right) \mathrm{d}t
%= - \lambda \int_0^1 \mathrm{d}t\,  {\rm Re}\left[{\partial \delta x\over \partial \lambda}\overline{\delta x} \right] 
=-{1\over 2} \lambda {\mathrm{d} (R^2) \over \mathrm{d} \lambda} \implies {\mathrm{d} h\over \mathrm{d}(R^2)}=-{1\over 2} \lambda.
\label{g1d}
\end{align}
Since $h=0$ at $R=0$, and $\lambda>\lambda_m$ on the $m$'th eigenflow, we infer that $h<-{1\over 2}\lambda_m R^2$, $\forall R$. 

\begin{figure}
\centering
\includegraphics[width = 0.5 \textwidth]{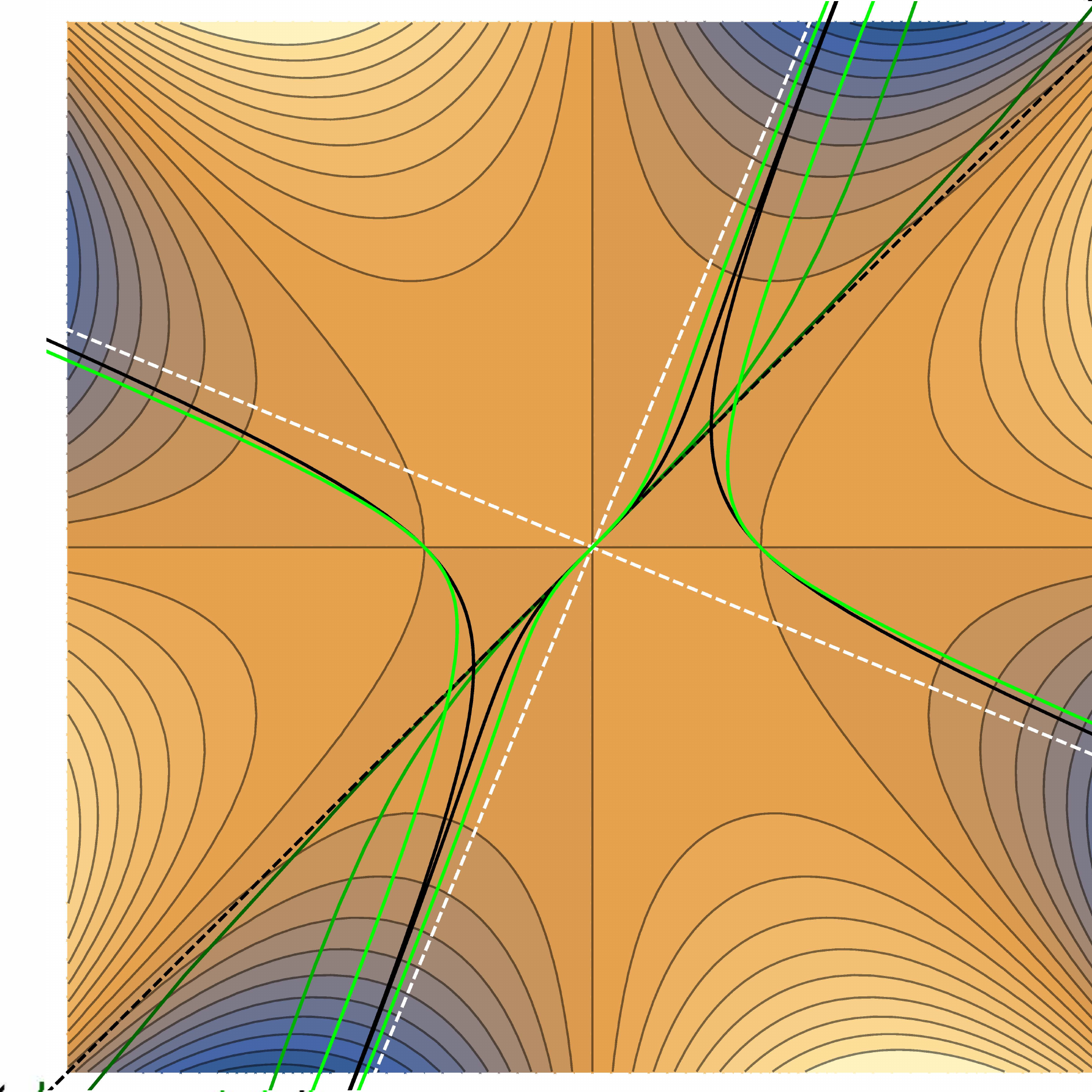}
\caption{Complex $x_m$-plane for the lowest mode ($m=1$) of the quartic oscillator $V(x)=x^2/2$, showing contours of $h$ calculated in the LFA approximation (see Section \ref{quartic} and Appendix \ref{etapp}). The original integration contour is the real axis. Curves show, again in the LFA approximation: a) the three eigenthimbles for the real saddles (light green), b) the steepest descent contours for the same saddles (black), c) the eigenthimbles for the trivial saddle at twice and four times the frequency (in dark and very dark green), d) the Wiener thimble ${\cal J}_0$ (black, dashed) and e) the steepest descent contours for the potential (white, dashed).  As $m$ is raised, the high frequency eigenthimble hugs the Wiener thimble out to larger and larger $|x_m|$, before bending towards a steepest descent contour for the potential. For explicit analytic formulae, see Appendix \ref{etapp}}
\label{fig:comparison}
\end{figure}

As we shall see shortly, instead of using $R$ it is actually better to use the intrinsic ``proper length" $L$ as the coordinate along any eigenflow. The proper length is given by the line element $\mathrm{d}L^2\equiv \int_0^1  |\mathrm{d} x(t)|^2\mathrm{d}t$, where $\mathrm{d} x(t)$ is the change in a complex path induced by an infinitesimal eigenflow, {\it i.e}, an infinitesimal increase in the Lagrance multiplier $\lambda$.  The change in the proper length $L$ along an eigenflow is generally greater than the change in $R$ because the eigenthimble is a curved manifold embedded in the (flat) Euclidean space of complexified paths which, formally, has twice the dimension. Provided the eigenthimble manifold is not {\it too} curved (as we expect, {\it e.g.}, from Fig.~\ref{fig:comparison}), we nevertheless expect $L^2$ to be in one to one correspondence with $R^2$. It is convenient to assign $L>0$ to $R>0$ and, likewise, $L<0$ to $R<0$, so that $L$ and $R$ agree when both are small.

Now consider an ordered succession of such flows, along the shallowest direction, the next shallow direction, and so on (see Fig.~\ref{fig:eigenthimble}). Every flow starts at a saddle of $h^+$, and is controlled by the Lagrange multiplier for a constraint on the ``radius squared" $R^2$ of the deviation of the path from that saddle. For example, after flowing a proper length $L_1$ down the shallowest direction, to a complex path $x_1(t)$, the second flow in Fig.~\ref{fig:eigenthimble} is obtained by extremizing $h^++{1\over 2} \lambda_2\int_0^1  (|x(t)-x_1(t)|^2-R_2^2)\mathrm{d}t$ for each $R_2$, and so on. With each successive eigenflow, the eigenvalues of the quadratic operator $-\hat{O}$, appearing in the kinetic term of the height function $h$, move upwards, equally, by the sum of the (positive) Lagrange multipliers $\lambda$ used so far. Thus, the eigenflows can {\it only} be followed in successively steeper directions. In order to fill out the entire high frequency eigenthimble, one must use such an ordered set of high frequency eigenflows, of proper lengths $L_{m},$ for $m\geq m_C$ (of both signs).  Since the frequencies $m$ are by assumption much higher than any involved in the saddle, we do not expect any obstructions to deforming the original, real integration domain of the high frequency mode amplitudes $x_m$ onto the Wiener thimble $\delta x_m= e^{i {\pi\over 4}} w_m $, with $w_m$ real, and then onto the closest steepest descent direction for the potential, at large $x_m$. 

\begin{figure}
\centering
\includegraphics[width = 0.5\textwidth]{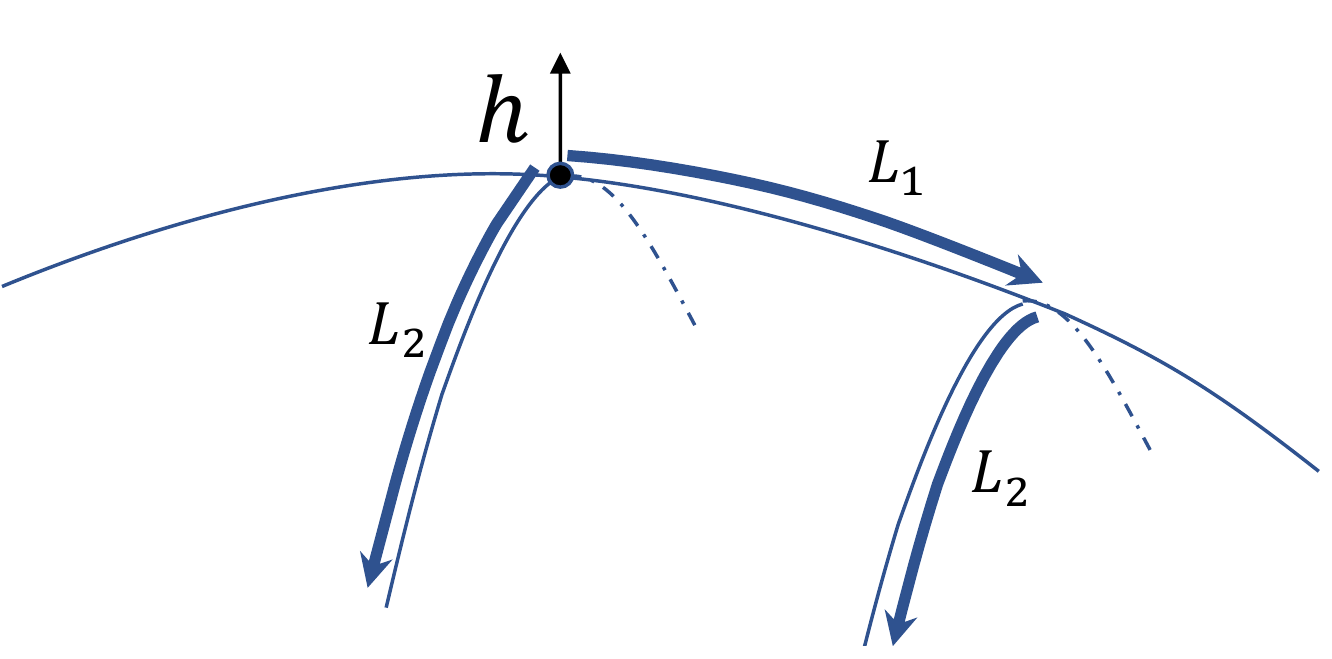}
\caption{Descent eigenflows in the height function $h$ in the shallowest direction, parameterized by the proper length $L_1$, the next shallowest, by $L_2$, and so on. Any point on the eigenthimble may be reached by a (correctly ordered) succession of such flows. }
\label{fig:eigenthimble}
\end{figure}

The reason that the proper length coordinates $L_m$, $m\geq m_C$ (which we shall collectively denote by ${\bm{L}}$) are useful is that they allow us to bound the magnitude of the complex Jacobian ${\rm det} (\partial x_m/\partial L_k) $ which appears as a factor in the integration measure when we deform every $x_m$ contour from the real $x_m$-axis onto the complex eigenthimble $x_m({\bm{L}})$. At {\it any} point on the eigenthimble, this Jacobian can be expressed as a magnitude times a phase factor, which contributes to the $e^{i \theta_C}$ in (\ref{e0}). The magnitude is bounded above by unity, just because the volume of a parallelipiped is bounded above by that of a cuboid with sides of the corresponding lengths. To see this, note that the eigenthimble is a curved manifold embedded in the complexified space of paths, whose natural metric is flat. The metric induced on the eigenthimble is given by $g_{kl}= {\rm Re}\sum_m\left[(\partial \overline{x}_m /\partial L_k)(\partial x_m /\partial L_l)\right]=\int_0^1  {\rm Re} \left[(\partial \overline{x}(t)/\partial L_k) (\partial x(t)/\partial L_l)\right]\mathrm{d}t$, expressed explicitly in terms of the complex paths $x(t)$. When integrating over the eigenthimble, we must therefore include a factor of $\sqrt{g}$ in the measure. But the determinant of a positive definite, symmetric matrix is bounded above by the product of its diagonal elements. In proper length coordinates, these diagonal entries are all unity. It follows that, in the proper length coordinates, the magnitude of the integrand for the path integral taken over the eigenthimble is bounded above by the simple quantity $e^{h/ \hbar}$. We shall now argue this is bounded by the corresponding expression for a free particle.

\section{Bounding $h$ on the high frequency eigenthimble} 
\label{eigenthimble}

We now wish to bound the height function $h\equiv {\rm Re}(i S[x])$ on the high frequency eigenthimble. As in the previous section, to keep the notation minimal, we shall refer to the fluctuation as $x(t)$ and the difference in the height function from its value at the saddle as $h$. We shall also assume the potential $V(x)$ is a polynomial which, for real $x$, is bounded below. As argued above, with increasing frequency, the interacting eigenthimble closely follows the free particle thimble, further and further out into the complex $x_m$-plane (see Fig.~\ref{fig:comparison}). Therefore, at high frequencies, significant deviations from the free particle thimble arise only at large $R^2$ where the integrand is very small. This greatly simplifies our analysis: at high frequencies, we can neglect contributions involving the classical solution or the boundary conditions and focus on the ``amplitude to go nowhere," ignoring all but the highest power in the potential. 

Consider the high-frequency, Gaussian measure defined by $\displaystyle  h_G=-{1\over 2} \sum_{m=m_C}^\infty \lambda_{m} L_m^2.$ As $m_C$ is raised, the contributing eigenvalues $\lambda_m$ approach those of the free particle $\lambda_{0,m}$, and $h_G$ tends to the high-frequency, free particle height function $h_0$.  Along any descent eigenflow, both $h$ and $h_0$ are negative and decrease monotonically. From (\ref{g1d}), their relative decrease along the $m$'th eigenflow is
\begin{align}
{\mathrm{d} h\over \mathrm{d}h_0} ={\lambda \over \lambda_{0,m}} {\mathrm{d}(R^2)\over \mathrm{d}(L^2)}
\label{g1dd}
\end{align}
For large $m_C$, the rhs tends to unity at small $L^2$ and to infinity at large $L^2$. In the next section, we shall calculate it explicitly for the quartic oscillator and show it exceeds unity for all $L$. We conclude that  $\mathrm{d}h/\mathrm{d}h_0 \geq 1$ and therefore that $h \leq h_0$ along any eigenflow.

With one more assumption, we can extend this  bound to the whole eigenthimble. We assume that the ``shallowest," ``next shallowest"  and so on directions down from the saddle are unique in the following sense.  Namely, the gradient of the height function with respect to each of the $L_m$, $m_C\leq m$,  is minimized when the coordinates on previously traversed eigenflows, $L_i$, $m_C\leq i<m,$, are all zero. In other words, for any change $\mathrm{d}L_m$, in any particular $L_m$, which increases the proper distance from the saddle, $\mathrm{d}L_m \partial_m h (L_{m_C},\dots, L_{m-1},L_m,{\bm{0}})\leq \mathrm{d}L_m \partial_m  h({\bm{0}},L_m,{\bm{0}})$. Now consider an infinitesimal change in {\it all} of the $L_m$ on the eigenthimble, each change again increasing the proper distance from the saddle. The corresponding decrement in $h$ is $\displaystyle \mathrm{d} h= \sum_{m=m_C}^\infty \mathrm{d}L_m \partial_m h(L_{m_C},\dots, L_{m-1} ,L_m,{\bm{0}}) $. By the uniqueness assumption, this is bounded above  by $\displaystyle \sum_{m=m_C}^\infty \mathrm{d}L_m \partial_m  h({\bm{0}},L_m,{\bm{0}})$.  Each term on the rhs is bounded by the corresponding free particle term, as argued in the paragraph above. Hence, $\mathrm{d} h \leq\mathrm{d}h_0$ everywhere on the eigenthimble, proving the result.

Once we have bounded the high frequency integrand for  the interacting theory by the corresponding integrand for the free theory, we may use Lesbesgue's dominated convergence theorem to establish that the path integral for the interacting theory leads to a well-defined measure (See Appendix \ref{measuretheory} for a discussion of infinite-dimensional measures). Writing the exponent in terms of its real and imaginary parts, $iS[x]= h[x]+iH[x]$, the path integral over high frequency modes, on any relevant thimble $\mathcal{J}$ may be written as $ \int_{\mathcal{J}}\left({\rm det} {\partial {\bm{x}}\over \partial {\bm{L}}}\right) e^{(h[x]+i H[x] )/\hbar} \prod_{m=m_C}^\infty \sqrt{\lambda_{0,m} \over 2 \pi i \hbar} \mathrm{d}L_m$. Recalling the bound on the magnitude of the Jacobian given at the end of Section \ref{eigenflow}, $| {\rm det}{\partial {\bm{x}}\over \partial {\bm{L}}}|\leq 1$, where the ${\bm{L}}$ are the ordered, proper length coordinates on the high frequency eigenthimble,  ${\bm{L}}=(L_{m_C},L_{m_C+1},\dots)$, we infer the following bound on high frequency path integral, defined in (\ref{g1b}):
\begin{align}
& \left|\int_{\mathcal{J}} \left({\rm det} {\partial {\bm{x}}\over \partial {\bm{L}}}\right) e^{(h^H[x]+i H^H[x] )/\hbar}   \prod_{m=m_C}^\infty  \sqrt{\lambda_{0,m} \over 2 \pi i \hbar}  \mathrm{d}L_m  \right|\leq  \int_{\mathcal{J}}  \left| {\rm det}{\partial {\bm{x}}\over \partial {\bm{L}}}\right| e^{h^H[{\bm{L}}]/\hbar} \prod_{m=m_C}^\infty   \sqrt{\lambda_{0,m} \over 2 \pi \hbar}\mathrm{d}L_m  \nonumber \\
&\leq \int_{\mathcal{J}} e^{(h^H[{\bm{L}}]-h^H_0[{\bm{L}}])/\hbar}  \left(e^{h^H_0[{\bm{L}}]/\hbar} \prod_{m=m_C}^\infty \mathrm{d}L_m \sqrt{\lambda_{0,m} \over 2 \pi \hbar}  \right) 
\equiv \int_{\mathcal{J}} g^H[{\bm{L}}]\mathrm{d}\mu_{B}^H,
\label{g9}
\end{align}
where, as before, the superscript $H$ denotes the high frequency part. In the last step we have identified $\displaystyle d\mu_{B}^H=\prod_{m=m_C}^\infty \left(  e^{-{(m \pi L_m)^2\over 2 W^2}  } {m \pi \over \sqrt{2 \pi} W}  \mathrm{d}L_m   \right)$, with $W^2=\hbar T/M$, as the high frequency part of the Brownian Bridge measure of weight $W$, defined in (\ref{g1w}) and discussed further in Appendix \ref{measuretheory}. Note that the Wiener coordinates $w_m$ used there  are, in fact, the proper lengths $L_m$ on the free particle thimble.  In (\ref{g9}) we have also defined $g^H[{\bm{L}}]\equiv e^{(h^H[{\bm{L}}]-h^H_0[{\bm{L}}])/\hbar}$ which is obviously positive.

In the next section, we shall show by explicit calculation that, for a polynomial potential with highest power $x^4$, $g^H({\bm{L}})$ is bounded above by unity. The calculation may be straightforwardly extended to any polynomial potential $V(x)$ which is bounded below: assuming the same result, the bound of Eq. (\ref{g9}) suffices to show that, with our definition using eigenthimbles, the interacting path integral yields a well-defined measure.

If we denote by $A$ a set of complex paths on the eigenthimble  ${\mathcal{J}}$, with each element $X$ of the set corresponding to some particular real coordinates $L$,  then the combination $g[X]\, \mathrm{d}\mu_{B}$, where $g[X]$, is positive and bounded (we henceforth drop the superscript $H$), defines a proper $\sigma$-measure $\nu(A) = \int_A g[X]\, \mathrm{d}\mu_{B}[X]$ on the space of paths on the thimble since, (i) $\nu(A)$ is non-negative, (ii) the measure of the empty set vanishes $\nu(\emptyset) = \int_{\emptyset} g[X]\mathrm{d} \mu_{B} [X] = 0$, and (iii) for a sequence of mutually disjoint sets $A_n$ and its union $A= \cup_{n=1}^\infty A_n$, the measure of the union coincides with the sum of the measure of the sets:

\begin{align}
\nu(A) = \int_A \mathrm{d}\mu_{B} \, g[X] = \lim_{N\to \infty} \int \mathrm{d}\mu_{B} \, g[X] \left( \sum_{n=1}^N 1_{A_n}\right)= \lim_{N\to \infty} \sum_{n=1}^N \int\mathrm{d}\mu_{B} \,g[X]  \,1_{A_n} = \sum_{n=1}^\infty \nu(A_n).
\label{g10}
\end{align}
Here, using the fact that $g[X]$, being positive and bounded, is an integrable function with respect to the measure $\mu_B$, we were able to exchange the summation and the integral, for all $N$, by the Lesbegue dominated convergence theorem (see Appendix \ref{theorems}). 

We have therefore succeeded in bounding the infinite-dimensional path integral over high frequency modes with a well-defined measure on the space of complex paths. In taking the absolute value of the integrand, we eliminate the complex phases arising from a) the complex Jacobian ${\rm det} (\partial x_k/\partial L_m)$ and b) $e^{i H/\hbar}$ where $H\equiv {\rm Im}\left[i S[ x(L)]\right]$. These two phases are combined to form the factor $e^{i\theta_C}$ appearing in Eq.~(\ref{e0}). This factor only improves the convergence of the integrals and hence does not change the conclusion.  Therefore, we have shown that the path integral defined by contour deformation exists as a well-defined complex measure. 

Having used eigenflow methods to establish the existence of real time path integrals, a natural next question is how useful might they be for practical calculations? We shall make a general comment here, before moving on to a detailed discussion of the quartic oscillator. We have defined a countable infinity of high-frequency, nonlinear eigenflows, $x(L_m,t)$, $m\geq m_C$, in terms of the proper lengths along these flows. The proper length coordinates $L_m$ are a natural, nonlinear generalization of the eigenmode amplitudes $x_m$, used to define the space of perturbations about the saddle. (Recall, the operator  $\hat{O}$ associated with quadratic fluctuations about the saddle has a complete set of eigenfunctions $\psi_m(t)$). One is therefore tempted to propose the following general ansatz to cover the whole eigenthimble: 
\begin{align}
x(L, t)=\sum_{m=1}^{m_C-1} x_m  \psi_m(t)+\sum_{m=m_C}^\infty x(L_m,t).
\label{g11}
\end{align}
A considerable simplification at high frequency, which we shall detail below for the quartic oscillator, is that the nonlinear amplitudes for the high frequencies take a universal form, $ x(L_m,t)=m x_1(L_m/m, m t)$, so that a single function $x_1(L_1,t)$ describes them all. In principle, we should now integrate out the high frequency $L_m$, over the real domain, to obtain the effective action for the remaining, low-frequency parameters $x_m$, with $m<m_C$. One should then identify the relevant saddles in $x_m$ and integrate over their associated steepest descent contours.

\section{the quartic oscillator}
\label{quartic}

In this section, we check these ideas in detail for the quartic oscillator, defined by the action $S=\int_0^1  {1\over 2} (T^{-1}\dot{x}^2-T x^4)\mathrm{d}t$. We have set the mass and quartic coupling to unity by rescaling $T$ and $x$. In this case, there is in fact a trivial contour rotation which makes the Lorentzian path integral converge. Namely, setting $x(t)=e^{i \pi/4} w(t)$ takes the exponent in the path integral to $i S/\hbar = (1/\hbar)\int_0^1  {1\over 2} (-T^{-1}\dot{w}^2+i T w^4)\mathrm{d}t$. The kinetic term is then that which leads to the Brownian bridge measure for the free particle and the potential term is a pure phase which only improves the convergence. Therefore, this transformation allows us to trivially bound the contour-rotated path integral by the Brownian bridge measure, showing that it exists, at least for ``going nowhere" boundary conditions which the contour rotation respects. While this simple argument is reassuring, it is also very limited since it applies to only one theory. We shall, in this section, develop an approach which can be extended to much more general theories, and use the quartic oscillator to illustrate it.  

Unlike the harmonic oscillator, for any initial and final values of $x$ and any time $T$, there are an infinite number of real, classical solutions. This is a generic property for any potential $V(x)$ rising faster than quadratically at large positive and negative $x$. Essentially, if you throw a ball faster, it will bounce more times off the ``walls" on either side, before it reaches its destination at some fixed, later time. 
Here we shall concentrate on the high frequency eigenthimble, where various simplifications occur. For one thing, we can concentrate on the amplitude ``to go nowhere," in which no scale other than the time $T$ enters the problem. The classical equation of motion is $\ddot{x}=-2\,T^2\, x^3$. The general solution satisfying $x_C(0)=0$, is $x_C(t)= C T^{-1} {\rm sn}(C t,-1)$, for an arbitrary constant $C$. Here, ${\rm sn}(u,m)$ is the Jacobi elliptic function of argument $u$, with parameter $m$, and $K(m)$ its quarter-period~\cite{Abramowitz}. The simple dependence on $C$ is a consequence of the classical scale symmetry. Jacobi elliptic functions are doubly periodic in the complex plane of their argument. Their zeros occur on a square lattice: for $m=-1$ and for real argument, zeros occur at integer multiples of $2 K(-1) \approx 2.622$.  The corresponding real, classical solutions, satisfying $x_C(1)=0$ are\footnote{There are also an infinite number of regular, complex classical solutions, satisfying $x_C(0)=x_C(1)=0$. These are found by setting $n=a+i b$ where $a, b\in \mathbb{Z}$. Up to equivalence, the nontrivial solutions have $a>0$, $|b|<a$. We exclude cases where $b/a$ is a ratio of odd integers  since then (\ref{ex2}) has a pole at some intermediate $0<t<1$.  The action for these solutions is $S_C={1\over 6} (2 K(-1) n)^4/ T^3$. Thus, for negative $b$ the height function $h={\rm Re}[i S]$ is positive and the saddles cannot be relevant. For positive $b$, $h$ is negative: by this criterion, the saddles are potentially relevant. However, as we show in detail in Ref.~\cite{FT2}, they are in a topological sector disconnected from the free particle and are irrelevant in the Picard-Lefschetz sense. }

\begin{align}
x_C(t) = \pm {2 K(-1)\over T} n \,{\rm sn} ( 2 K(-1) n t,-1), \quad n\in \mathbb{Z}_+.
\label{ex2}
\end{align}
Being real, these saddles are all relevant so the sum over saddles in formula (\ref{e0}) is infinite and must be carefully regulated. As a result of this infinite sum, the quantum propagator is a {\it distribution}, not a function (this is nicely emphasized, for example, in~\cite{Fulling:2003}). In a companion paper, we generalize these solutions to the case of arbitrary initial and final conditions, and compare the propagator calculated as a sum over saddles with the exact propagator calculated by solving the time-dependent Schr\"odinger equation.

\begin{figure}
\centering
\includegraphics[width = 0.4 \textwidth]{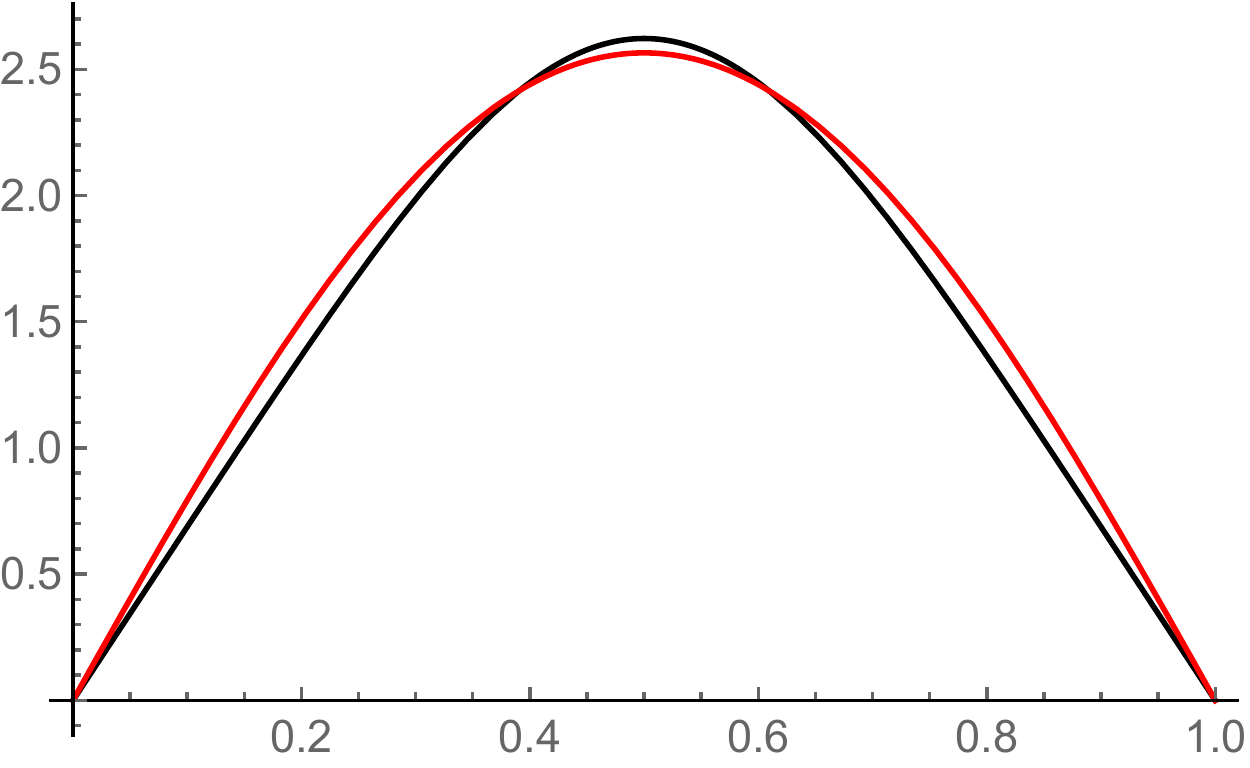}
\caption{For the amplitude to ``go nowhere" in the quartic oscillator, {\it i.e.}, the boundary conditions $x(0)=x(1)=0$, we compare the exact, analytic solution (black curve) to the LFA approximation (red curve), for $T=1$. The energy and action obtained in the LFA exceed the exact results by just over 3 per cent.}
\label{fig:compare1}
\end{figure}

The classical solutions (\ref{ex2}) may be represented as a rapidly convergent Fourier series. In this section, we shall make use of a simple approximation we call the ``lowest frequency approximation" or LFA. Starting from a given Fourier mode satisfying the boundary conditions, for example, $\sin m \pi t$, with $m$ an integer, the  nonlinear term in the equation of motion sources higher Fourier modes. When an odd power of a Fourier mode is expressed in Fourier modes the term with the largest coefficient is the lowest frequency term, involving the original mode (essentially as a consequence of the structure of Pascal's triangle). Higher Fourier modes are sourced with decreasing amplitude. Lambert's expansion of Jacobi elliptic function shows this explicitly~\cite{Abramowitz}. It is therefore a reasonable starting approximation to include only the lowest Fourier mode. To see how well the LFA reproduces the classical solutions (\ref{ex2}), we simply set $x(t)=x_1 T^{-1} \sin (\pi t)$ and ignore higher modes.  The action then reads $S_{LFA}={1\over 4 T^3}(\pi^2 x_1^2 -{3\over 4} x_1^4)$. As well as the trivial saddle at $x_1=0$, this has two real saddles at $x_1=\pm \sqrt{2\over 3} \pi$. Both are clearly relevant. The exact solution has action $S={8\over 3} K(-1)^4 T^{-3}\approx 7.878 \,T^{-3}$ whereas the LFA approximation yields $S={1\over 12} \pi^4T^{-3}\approx 8.117 \,T^{-3}$, just over 3 per cent higher. The energies differ in the same proportion. Fig.~\ref{fig:compare1} compares the exact solution with the LFA approximation. 

\newpage\noindent
We now turn to the main purpose of this section, which is to calculate the high frequency eigenthimble and show the height function $h={\rm Re}\left[i S[x]\right]$ is bounded by that for the free particle. As discussed in previous sections, at high frequencies the initial and final conditions become irrelevant and we can focus on the amplitude ``to go nowhere" and on the trivial saddle $x_C(t)=0$.  For simplicity we set $T$=1. We start by rotating the kinetic term, setting $\delta x(t)=e^{i \pi/4} w(t)$ to obtain $h={1\over 4} \int\left(-\dot{w}^2-\dot{\overline{w}}^2+i (w^4-\overline{w}^4\right) \mathrm{d}t$. The eigenflow equation is $2 \delta h^+/\delta w=\ddot{w}+2 i w^3+\lambda \overline{w}=0$. Setting $w=F+iG$, the real and imaginary parts are 
\begin{align}
-\ddot{F}-\lambda F=2 G^3-6 F^2 G, \qquad -\ddot{G}+\lambda G=2 F^3-6 FG^2.
\label{g1ci}
\end{align}

We shall solve these equations  in three ways. Near the saddle at $F=G=0$, the kinetic terms dominate. We therefore use perturbation theory in the potential, {\it i.e.}, the nonlinear terms in (\ref{g1ci}). In the opposite limit, the potential dominates and the solution runs out along a steepest descent direction for the potential. In these two, opposite regimes, we can solve (\ref{g1ci}) analytically.  We connect them with the ``lowest frequency approximation" or LFA. As explained above, this approximation reduces odes like the classical equations of motion, or (\ref{g1ci}), to algebraic equations. Although only an approximation, it is remarkably accurate even at lowest order, agreeing well with the first and second treatments in their regions of validity. It can also be improved systematically by including more and more Fourier modes and using a numerical root-finding algorithm in which the LFA provides a starting point for the search. We shall show, using the LFA and its systematic improvement, that the height function $h$, along any eigenflow and for all proper lengths $L$,  is lower than the free particle height function $h_0$ at the same $L$. We shall also compare the two height functions at generic points on the eigenthimble, using the ansatz (\ref{g11}), finding that $h$ is bounded above by $h_0$, everywhere. 

Let us start with perturbation theory in the potential. To zeroth order in the potential, the lhs of the first equation in (\ref{g1ci}) has an infinite number of solutions, $F_0= A \sin (m \pi t)$, $\lambda = \lambda_0\equiv (m \pi)^2$ with $m$ a positive integer and $A$ an arbitrary constant. Each of these solutions ``seeds" an eigenflow, {\it i.e.}, a nonlinear solution which may be found as a power series in $A$. The operator on the lhs of the second equation in  (\ref{g1ci}) is positive definite and so is invertible. At leading order, we find $G_1\sim A^3/\lambda_{0,m}$. Then, from the first equation, ignoring any new contribution to the seed mode zero mode of the linear operator, $F_1\sim F_0^2 G_1\sim A^5/\lambda_0^2$, and so on.  Hence, the nonlinear solution is obtained as a power series in  $A^2/\lambda_0$. 

Explicitly, for each $m$ we find
\begin{align}
F(t)&=A \left(\sin (m \pi t) +{A^4\over 320 \lambda_{0,m}^2} \left(51\sin (3 m \pi t) - \sin (5 m \pi t)\right)+\dots \right); \cr  G(t)&=A\left({A^2\over 20 \lambda_{0,m}} \left(15 \sin (m \pi t) - \sin (3 m \pi t)\right)+\dots\right),
\label{g1cii}
\end{align}
from which we obtain
\begin{align}
A&=\sqrt{2} R\left(1-{113\over 100} {R^4\over  \lambda_{0,m}^2} +\dots\right);\quad
\lambda=\lambda_{0,m} \left(1+{69\over 5} {R^4\over  \lambda_{0,m}^2} +\dots\right);\cr
L&=R\left(1+{113\over 125} {R^4\over  \lambda_{0,m}^2} +\dots\right);\quad
\delta h=\delta h_0\left( 1+{349\over 125}  {L^4\over  \lambda_{0,m}^2} +\dots\right),
\label{g1cii}
\end{align}
where $\delta h_0=-{1\over 2} \lambda_{0,m} L^2$, as above. The last result implies $\delta h/\delta h_0>1$ so that, along any eigenflow, parameterized by the proper length $L$ from the saddle, the interacting height function is bounded above by the free particle height function, at least when perturbation theory in the potential is valid.

\begin{figure}
\centering
\includegraphics[width =  \textwidth]{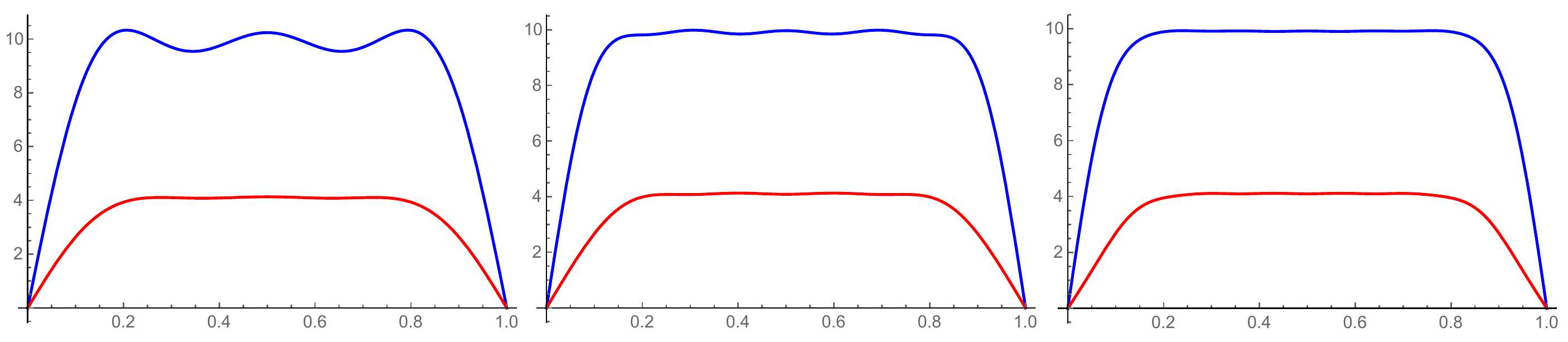}
\caption{Numerical calculation of the eigenthimble: the complex path is shown in Wiener coordinates, $x_m=e^{i \pi/4}w_m$, with $w_m=F+i G$ ($F$ is shown in blue and $G$ in red). The $m=1$ eigenflow solution  to (\ref{g1ci}) is shown, for $\lambda=230.09$ and using $N=3,5$ and $7$ modes (left to right), showing good convergence.}
\label{fig:eigenthimble}
\end{figure}

The series expansions given above hold for $L^2< \lambda_{0,m}$.  In the opposite regime, $L^2> \lambda_{0,m}$, over most of the solution the potential dominates and the complex path follows a direction of steepest descent for the potential, in the complex $x$-plane. In this approximation, we can again find the solutions analytically. For a quartic potential, the steepest descent direction, closest to the free thimble is $x(t)\approx e^{i 3 \pi/8}X(t) $ with $X(t)$ real. Substituting $w(t)=e^{i \pi/8}(X(t)+i Y(t))$ into the eigenflow equation, we find $Y\approx \ddot{X}/(6\sqrt{2} X^2)$ at large $X^2$. Thus, at fixed frequency, $Y(t)$ tends to zero as $X(t)$ grows large. Neglecting $Y(t)$, the equation for $X$ becomes $-\ddot{X} +2 \sqrt{2}X^3 =\sqrt{2} \lambda X$, which we can solve analytically:
 \begin{align}
&\qquad \qquad X(t)=2^{3\over 4} n \sqrt{\mu} K {\rm sn} (2 n K t,\mu),\quad \lambda=2 \sqrt{2} n^2 (1+\mu)K^2,\cr
&R^2=2 \sqrt{2} n^2 K(K- E),\quad h=-{4\over 3} n^4K^3\left((1+2 \mu) K-(1+\mu) E\right).
\label{g1e}
\end{align}
 where ${\rm sn} (2 n K t,\mu)$ is a Jacobi elliptic function with parameter $\mu$ and $K\equiv K(\mu)$ and $E\equiv E(\mu)$ are the corresponding complete elliptic integrals~\cite{Abramowitz}. As $\mu$ tends to zero, both  $E$ and $K$ tend to ${\pi\over 2}$, $R$ tends to zero, $\lambda$ tends to $ \lambda_{0,m}$ and $X_n(t)$ tends to $R \sqrt{2} \sin (n \pi t) $. As $\mu$ tends to unity, $E$ tends to unity but $K$ diverges as $\sim{1\over 2} \ln (16/(1-\mu))$ so, from (\ref{g1e}), $\lambda\approx 2 R^2$. In this limit, $X(t)$ approaches a ``square wave," with $n-1$ zeros, and $h$ tends to $-{1\over 2} R^4 $, exactly as expected from (\ref{g1d}). We have explicitly described the eigenthimble both at small and large $R$. The former description holds for $R^2 <\lambda_{0,n}$ and the latter for $R^2 >\lambda_{0,n}$. In both limits, the height function for the interacting theory is bounded by that for the free theory, consistent with our general arguments.

\begin{figure}
\centering
\includegraphics[width = \textwidth]{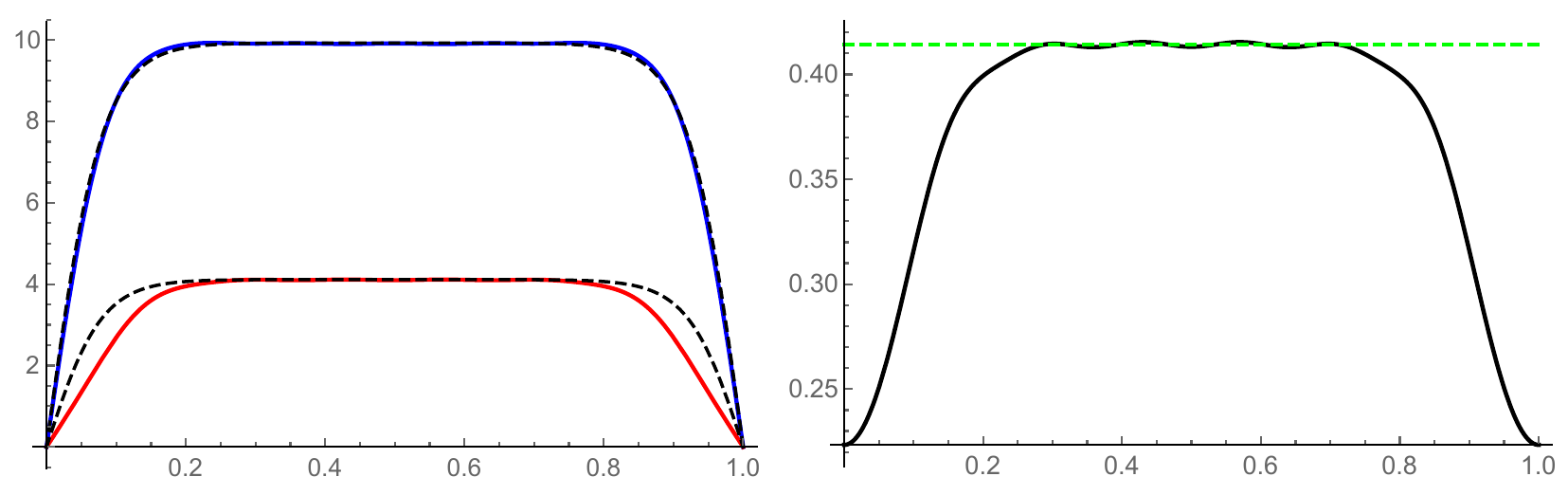}
\caption{(Left) Numerical calculation of the eigenthimble at $\lambda=230.09$, with $N=7$ modes included, $F$ in blue, $G$ in red, compared to the large-amplitude analytic approximation given in (\ref{g1e}), with $\mu=0.999955$ (dashed black lines in both cases), showing good agreement. (Right) The ratio $G(t)/F(t)$ is shown for the numerical solution with $N=7$ modes included. The green dashed line shows ${\rm tan}(\pi/8)\approx 0.4142$ which is the ratio predicted along the steepest descent contour for the potential. Both plots show that, unsurprisingly, the large amplitude approximation fails to provide an accurate description near the zeros of $F$ and $G$.  }
\label{fig:eigenthimble2}
\end{figure}

Next, we turn to a numerical solution of  (\ref{g1ci}), within an expansion about the LFA. We start from two observations. First, the scale symmetry of these equations means that all eigenmodes are expressible in terms of the lowest frequency solution: $F_n(t)=n F_1(n t), G_n(t)=n G_1(n t)$ and $\lambda_n= n^2 \lambda_1$. Second, the ``seed" frequency $m$ drives modes with frequencies which are odd multiples of $m$, but whose relative amplitude falls as the frequency rises. This is evident, for example, from the perturbative solution (\ref{g1cii}). The basic reason for this decline is the fact, easily checked, that the largest term in the Fourier expansion of $\sin (m \pi t)$, raised to an odd positive power, is the coefficient of $\sin (m \pi t)$ itself. Hence, if we set $\displaystyle w[t]=\sum_{m=1}^N (F_n+i G_n) \sin ( (2 n+1) m \pi t)$, with $F_n$ and $G_n$ real, we can expect convergence with increasing $N$. With this truncation, the height function $h^+$ is a polynomial in the mode amplitudes and the eigenflow equations (\ref{g1ca}) become {\it algebraic} equations, easily solved numerically with a root finding algorithm.  

\begin{figure}
\centering
\includegraphics[width = 0.6 \textwidth]{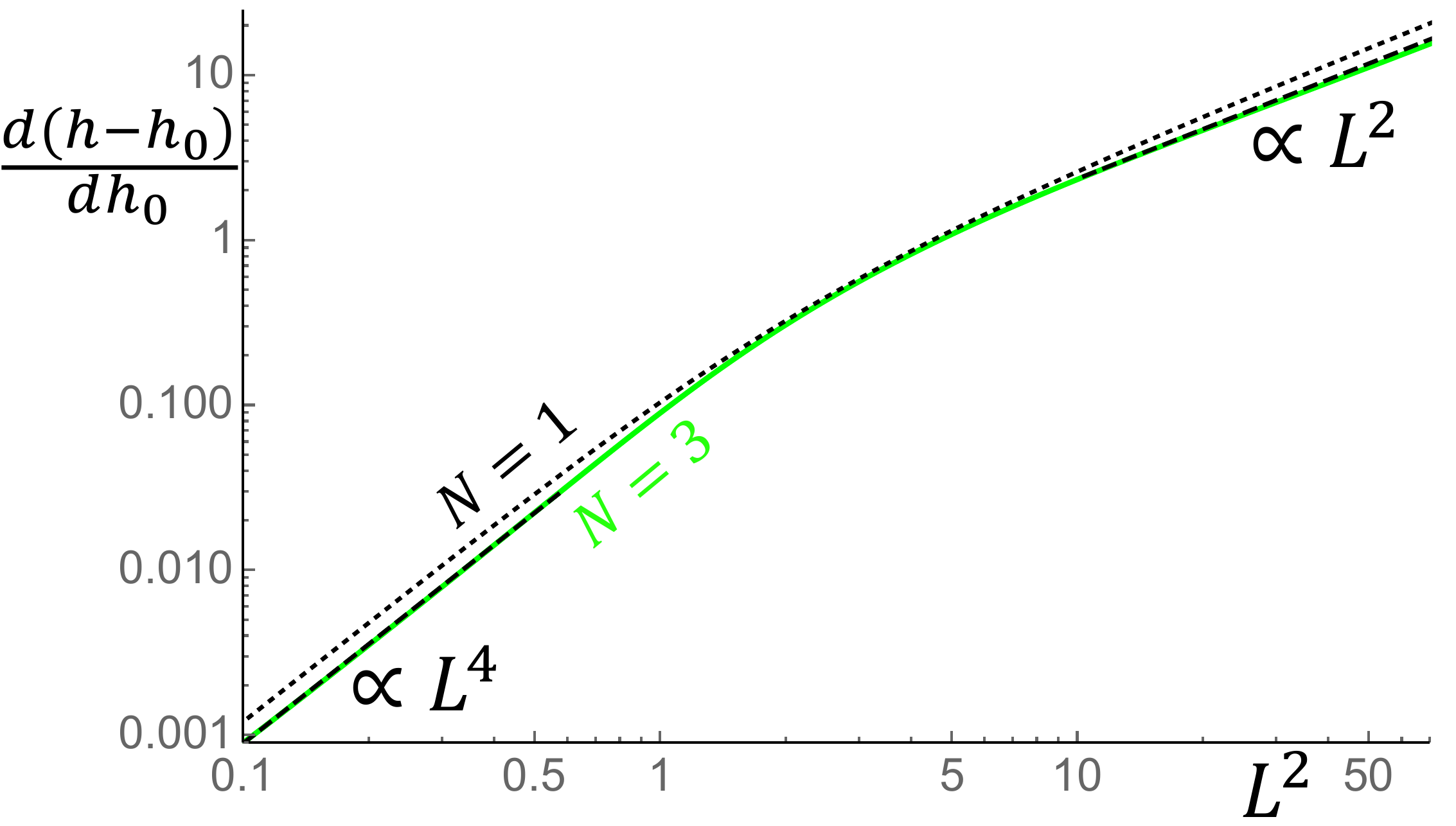}
\caption{The height function $h$ on the eigenthimble for the trivial saddle in the quartic oscillator, compared to that for the free particle, $h_0$, for all frequencies. The analytic approximation (\ref{g1g}) for $N=1$ (dotted black) is compared to the numerical result for $N=3$ (solid green). Dashed lines show analytic approximations at small and large $L$ (see text).}
\label{fig:lfa}
\end{figure}

For the quartic oscillator, we obtain
\begin{align}
h_{LFA}^+&=-{1\over 4} \lambda_0 (F_1^2-G_1^2)+{3\over 4}( F_1 G_1^3-F_1^3G_1)+{1\over 4}\lambda(F_1^2+G_1^2-2 R^2),
\label{g1f}
\end{align}
where $\lambda_0=(n \pi)^2$. The eigenflow equations, $(\partial   h_{LFA}^+/\partial  F_1)=(\partial  h_{LFA}^+/\partial  G_1)=0$ are solved by writing $F_1$ and $G_1$ in polar coordinates: $F_1=\sqrt{2} R \cos \theta$ and $G_1=\sqrt{2} R \sin \theta$. We find 
\begin{align}
R= \pm \sqrt{ \lambda_0\sin 2\theta\over 3 \cos 4\theta};  \, \, {\mathrm{d}L\over \mathrm{d}\theta} =\sqrt{\lambda_0 (3-\cos 8\theta) \over 6 \sin 2\theta (\cos 4 \theta)^3};\, \,
\lambda= {\lambda_0 \cos 2\theta \over \cos 4\theta}; \, \, h=-{\lambda_0^2 (\cos 2 \theta)^2 \sin 2\theta \over 6 (\cos 4\theta)^2}, 
\label{g1g}
\end{align}
for $0<\theta <{\pi\over 8}$, where the proper length $L$ along the flow is calculated from $\mathrm{d}L^2=\mathrm{d}R^2+R^2 \mathrm{d}\theta^2$. The second equation in (\ref{g1g}) is easily integrated numerically to find $L(\theta)$. The beauty of the LFA is that it is readily improved by adding more modes.  The height function $h^+$ is a polynomial in the mode coefficients. Including $N$ modes, the eigenflow equations $(\partial   h_{LFA}^+/\partial  F_n)=(\partial  h_{LFA}^+/\partial  G_n)=0$ provide $2 N$ equations for the $2N$ unknowns, $\{F_n,G_n\}$. The LFA analytic solution (\ref{g1g}) provides an excellent first approximation. With $\lambda$ held fixed, a standard root finding algorithm efficiently finds the solution (see Fig.~\ref{fig:lfa}). The solution converges rapidly with $N$: the result for $N=4$ is barely distinguishable from that for $N=3$.). 

\begin{figure}
\centering
\includegraphics[width = 0.9 \textwidth]{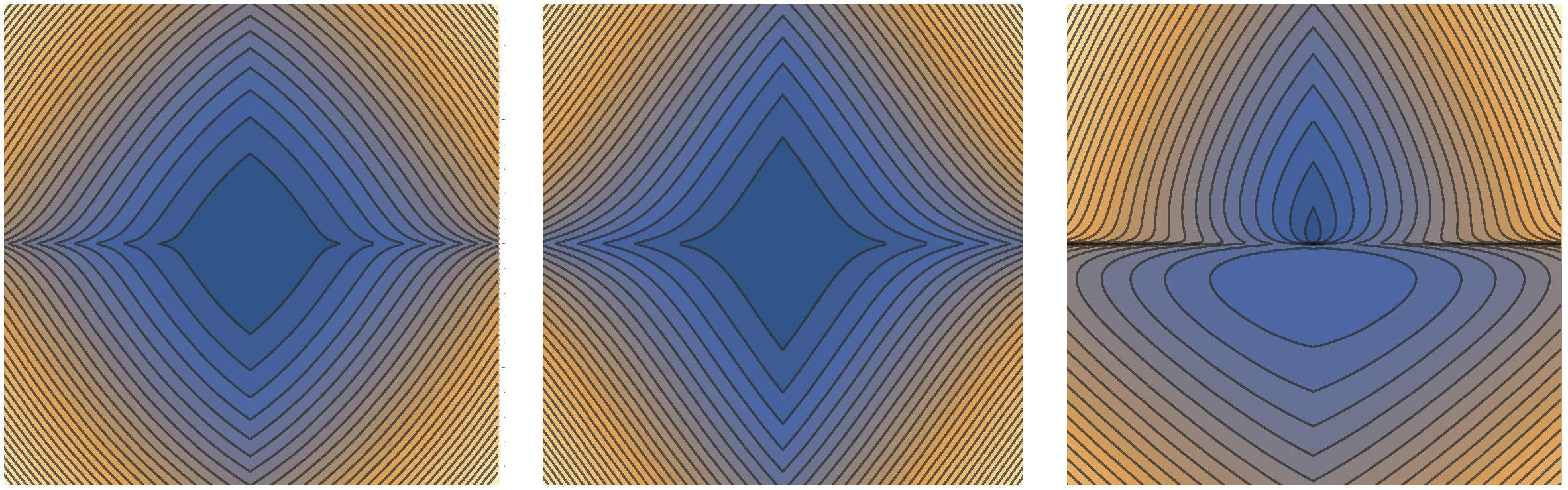}
\caption{For the amplitude to ``go nowhere" in the quartic oscillator, and for the trivial saddle $x(t)=0$, we show slices through the eigenthimble, as computed using the ansatz (\ref{g11}) with $m_C=1$,  using  the LFA approximation for the lowest 3 Fourier modes. We use the analytic formulae for the embedding coordinates of the eigenthimble, given for each frequency $m=1,2,3$ in terms of $\theta_m$, in (\ref{et3}) of Appendix \ref{etapp}. We integrate the formula (\ref{g1g}) for the $\mathrm{d}L_m/\mathrm{d}\theta_m$ to determine the corresponding proper lengths, which are added in quadrature to obtain the height function for the corresponding free particle. The plots show contours of $h/h_0-1$ against two of the three $\theta_m$ coordinates, where $h_0$ is the height function for the free particle. Writing $(\theta_1,\theta_2,\theta_3)={1\over 8}{\pi}(x_1,x_2,x_3)$, the left plot shows the region $-0.2<x_1,x_2<0.2$, at $x_3=0$, the middle plot shows the region $-0.2<x_2,x_3<0.2$, at $x_1=0$ and the right plot shows the region $-0.2<x_2,x_3<0.2$, at $x_1=0.2$. All contours are positive, with the contour heights growing from $0.001, 0.0014$ and $0.004$  at the centre of the plots, to $0.03, 0.04$ and $0.07$ at the boundaries, for the left, middle and right plots respectively. The free particle height function $h_0$, evaluated at $\theta_n=0.2$, with the other $\theta$'s zero, is $\approx -9.3 \,n^2$ for $n=1,2,3$.}
\label{fig:trivialthimble}
\end{figure}

Having shown the LFA to be a reasonable first approximation to each of the exact eigenflows, it is natural to ask whether the linear ansatz (\ref{g11}) suffices to bound the entire eigenthimble by that for a free particle. For the eigenthimble corresponding to the trivial saddle $x(t)=0$, and using the LFA, we find that (\ref{g11}) with $m_C=1$ gives an approximation to the eigenthimble for which the height function is already bounded above everywhere by that of the free particle. The analytical expressions for the eigenthimble in the LFA are given in Appendix \ref{etapp}.

In Fig.~\ref{fig:trivialthimble}, we plot contours of the quantity $h/h_0-1$ on the eigenthimble, where the free particle height function is calculated at the proper length ${\bm{L}}=(L_1,L_2,L_3)$, with $L_m$ is the length associated with the eigenflow along the Fourier mode $\sin m \pi t$. The three plots show three two-dimensional slices in the three corresponding coordinates  $\theta_m$, $m=1,2,3$, which parameterize the eigenflow curves in the complex $x_m$-planes. For all slices examined, no negative values occurred, indicating that the height function for the interacting theory, $h$, is bounded by that for the free theory, $h_0$, even at this lowest order of approximation.

 \section{Advantages of the real time path integral}
 \label{advantages}
 
 The real time path integral is considerably more intricate than the Euclidean one. However, it possesses some clear advantages, both conceptual and practical. 
 
Its first virtue is that it exists in some interesting cases where the Euclidean path integral does not. As we mentioned in the introduction, our main motivation is quantum gravity and cosmology, where the non-positivity of the Euclidean Einstein-Hilbert action for gravity (the well-known conformal factor problem as well as the lapse function problem for a de Sitter-like universe, highlighted on p.1 of Ref.~\cite{Feldbrugge:2017}) presents a major obstacle. Other very important contexts where standard Euclidean methods fail, even for equilibrium systems, include quantum matter in the presence of a chemical potential or a magnetic field, where the Euclidean action is complex and the path integral becomes oscillatory so its convergence may be problematic.  
 
A much more elementary example is the inverted harmonic oscillator (IHO), which is a useful toy model for a variety of physical phenomena~\cite{Subramanyan:2021}. Since the potential is unbounded below, the Euclidean path integral does not exist (see, for example the nice discussion in Ref.~\cite{Carreau:1990}). However, there is no problem with the quantum dynamics, as described by the time-dependent Schr\"odinger equation \cite{Wheeler:1959, Barton:1986}. Although the model is classically unstable, unstable trajectories only grow exponentially and so do not reach infinity in finite time. Furthermore, the semiclassical approximation becomes increasingly accurate the farther out the particle is from the origin. As a consequence of both facts, the evolution of wavepackets is well-defined and the total probability is conserved.

 For Gaussian actions, the real time path integral is easily performed, revealing interesting differences between the simple harmonic oscillator (SHO) and the IHO, for example. Their respective actions are $S_\pm[x]= \int_0^1  {1\over 2} M \left(\dot{x}^2/T  \mp T\omega^2 x^2\right)\mathrm{d}t$. Since the equations of motion are linear,  there is a unique saddle $x_C(t)$ for any boundary condition $x(0)=x_0$ and $x(1)=x_1$, and $T>0$. The classical action yields the well-known semi-classical phase factor $e^{i S/\hbar}$ in the propagator. To obtain the prefactor, we integrate over the fluctuations $\delta x(t)$.  The hessian operator $\hat{O}_\pm$ has eigenfunctions $\sqrt{2} \sin (m \pi t)$ with eigenvalues $\lambda_{\pm,m}= M \left((m \pi)^2/T \mp \omega^2 T\right)$, for $m$ a positive integer. The fluctuation action is diagonal in this basis, $S_{C,f}={1\over 2} \sum_{m=1}^M \lambda_{\pm,m} \delta x_m^2$ and the integrals over the $\delta x_m$ decouple. 

For the IHO and the free particle, all eigenvalues are positive. The steepest descent thimble (which is identical to the eigenthimble, for quadratic actions) is the Wiener thimble $\delta x_m=e^{i \pi/4} w_m$ with $w_m$ real, for all $m>0$. The rotation from $x_m$ to $w_m$ cancels the $1/\sqrt{i}$ factor in (\ref{g1a2}). Performing the Gaussian integral over the $w_m$, the prefactor becomes 
 \begin{align}
\sqrt{M\over 2 \pi i \hbar T}  \prod_{m=1}^\infty \sqrt{\lambda_{0,m}\over \lambda_{-,m}}=\sqrt{M\over 2 \pi i \hbar T} \sqrt{\omega T\over {\rm sinh} (\omega T)}=\sqrt{M \omega \over 2 \pi i \hbar \,{\rm sinh} (\omega T)},
\label{ihop}
\end{align}
where the free particle eigenvalue $\lambda_{0,m}=M (m \pi)^2/T$ and we used a standard infinite product formula. The free particle result is obtained in the limit $\omega\rightarrow 0$, in which the potential is turned off. 
  
The SHO is a little more involved. Assume first that $\omega T/\pi$ is not an integer. Then all modes with $m<\omega T/\pi$ have {\it negative} eigenvalues. Hence, for these modes the descent thimble is $\delta x_m=e^{-i \pi/4} w_m$ with real $w_m$. For modes with $m>\omega T/\pi$ the descent thimble is $\delta x_m=e^{i \pi/4} w_m$ as before. Hence, from  (\ref{g1a2}),  relative to the free particle we find an overall phase factor of $e^{-i K \pi/2}$, where $K\equiv {\rm int} (\omega T/\pi)$ is the number of negative eigenvalues of $\hat{O}$. This contribution to the overall phase factor is well-known in semi-classical quantum mechanics: $K$ is the Maslow index~\cite{Maslow,Keller,Tannor:2007}. Integrating over the $w_m$ yields for the infinite product $\displaystyle \prod_{m=1}^\infty \sqrt{\lambda_{0,m}/ |\lambda_{+,m}|}= \sqrt{|\omega T/{\rm sin} (\omega T)|}$ which reproduces the standard result~\cite{Feynman:1965}. When $\omega T/\pi$ {\it is} an integer, the propagator diverges. The explanation is interesting: the simple harmonic oscillator is a degenerate case because the action is exactly quadratic. So, when a saddle degenerates (meaning that the hessian operator $\hat{O}$ develops a zero eigenvalue), because there are no terms in the exponent at higher order than quadratic, the locally flat eigendirection becomes exactly flat. Therefore, even at finite $\hbar$, the oscillatory integral diverges. In more generic situations, such as we describe below, there {\it are} higher order terms. There is a higher order saddle, and the integral is large but still convergent.

The real time path integral for the IHO is thus simpler than that for the SHO. For the Euclidean path integral, however, we must use the Euclidean action, obtained by setting $T=-i T_E$, with $T_E>0$ and $i S=-S_E$. One obtains $S_{-,E}[x]=  \int_0^1  {1\over 2} \left( \dot{x}^2/T_E  - T_E\omega^2 x^2\right)\mathrm{d}t$. Thus $S_E$ is unbounded below for modes with $m< \omega T_E/\pi$. It follows that, for Euclidean times $T_E>\pi/\omega$, the Euclidean path integral does not exist~\cite{Carreau:1990}. 

\begin{figure}
\centering
\begin{subfigure}[b]{0.3\textwidth}
\includegraphics[width=\textwidth] {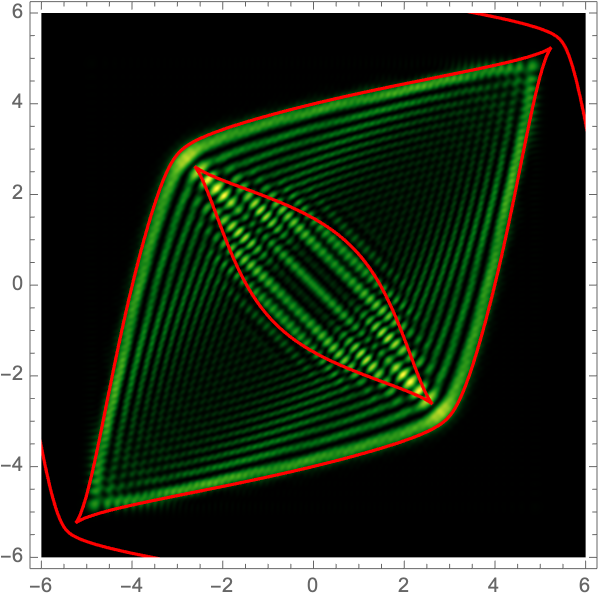}
\caption{Sum of eigenfunctions}
\end{subfigure}~
\begin{subfigure}[b]{0.3\textwidth}
\includegraphics[width=\textwidth] {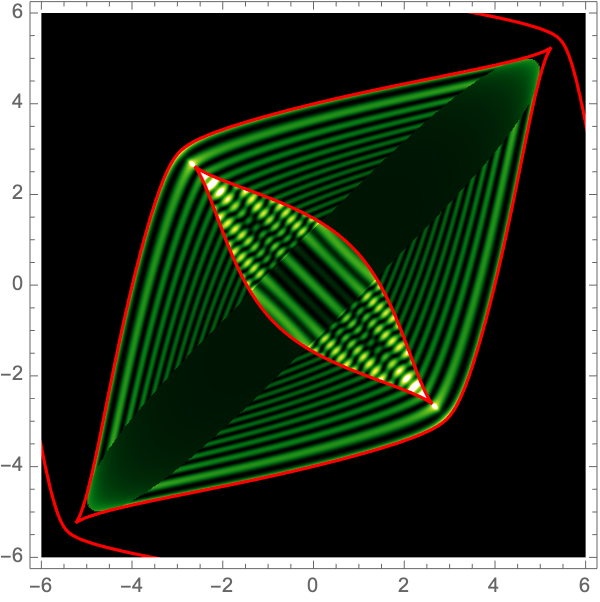}
\caption{Sum of saddles}
\end{subfigure}~
\begin{subfigure}[b]{0.3\textwidth}
\includegraphics[width=\textwidth] {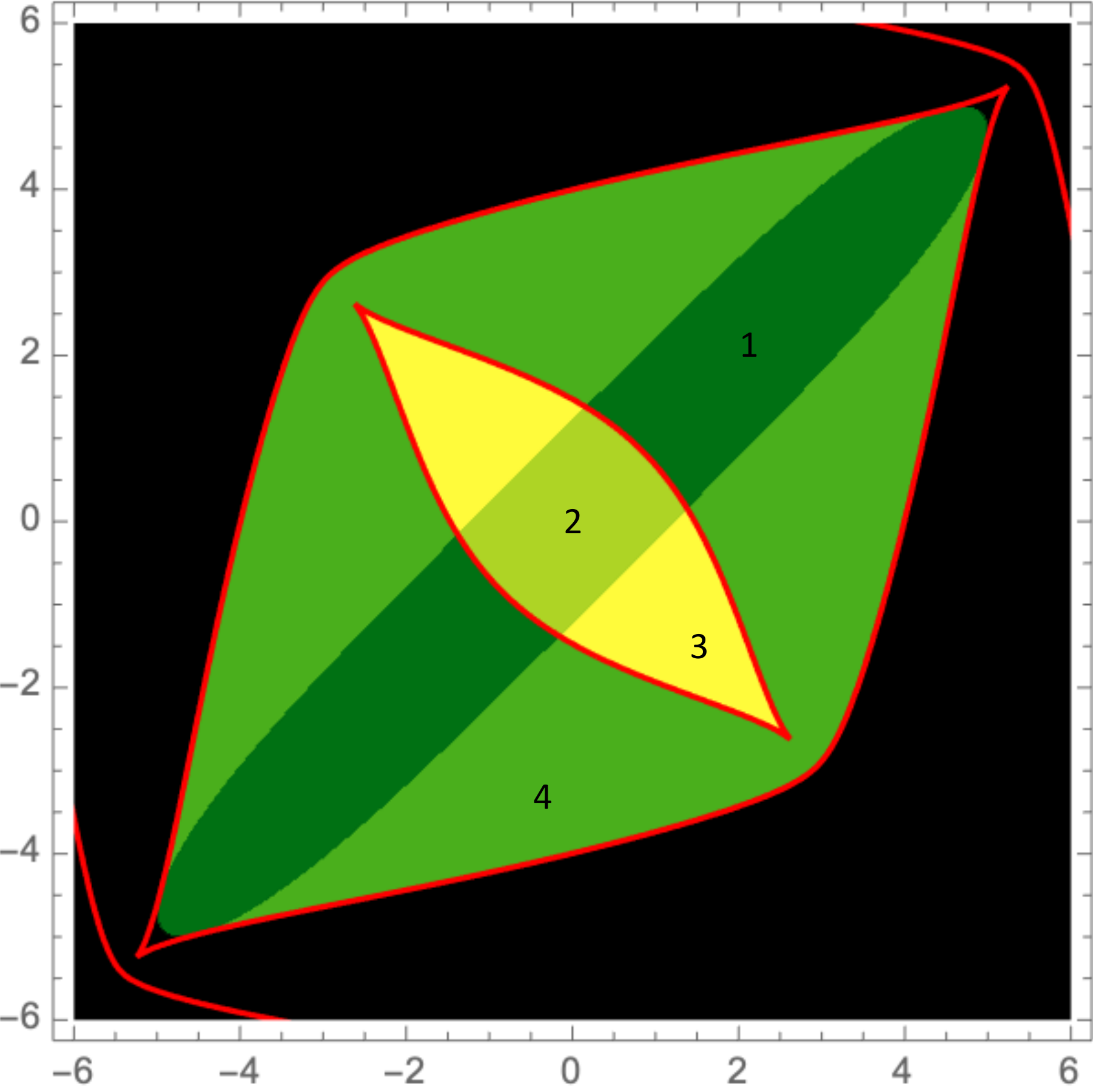}
\caption{Number of saddles}
\end{subfigure}~
\caption{The probability to propagate in unit time from an initial position $x_i$, shown on the horizontal axis, to a final position $x_f$, shown on the vertical axis, in the quartic oscillator $S={1\over 2} \int (\dot{x}^2-x^4)$, with $\hbar=1$. Classical ``fold" caustics are shown in red, ending in sharp ``cusps." (a) shows the traditional eigenfunction expansion employing the first $70$ energy levels; (b) shows the analytic result of the Picard-Lefschetz sum over real classical saddles, with the same energy constraint and (c) shows the number of such saddles at each value of $(x_i,x_f)$. The exact Picard-Lefschetz approach dramatically compresses the information required to accurately reproduce the pattern. Subtle differences between (a) and (b) are highly instructive (see text).}
\label{fig:quantumint}
\end{figure}

As this discussion illustrates, the real time path integral captures interference effects more directly than the Euclidean path integral, without any need for analytic continuation. Fig.~\ref{fig:quantumint} shows the classical caustics and the associated quantum diffraction pattern in the real time propagator for the quartic oscillator, calculated at leading semiclassical order. In the left panel, displaying the exact solution of the Schr\"odinger equation using the first $70$ energy eigenstates, we see that the probability to propagate from an initial position $x_0$ to a final position $x_1$ in real time $T=1$, consists of an intricate interference pattern governed by the caustics of the classical dynamical system. Note that a caustic is a classical phenomenon classified by Lagrangian catastrophe theory. It is defined as a configuration of the dynamical system for which one or more real, classical trajectories coincide~\cite{1978mmcm.book.....A}. (It is worth noting here the close correspondence between real time quantum physics and lensing in wave optics \cite{Feldbrugge:2019arXiv190904632F, 2022arXiv220412004J}). A small perturbation of the system either makes one or more classical solutions ``disappear" (or, more accurately, move into the space of complex classical trajectories) or split up into multiple real classical trajectories. In the central panel, we plot the semiclassical approximation of the corresponding path integral including only real trajectories for which the energy is below the $70$th energy level. There exists a close correspondence between the exact energy-truncated propagator and the semiclassical approximation. In the right panel, we indicate how the number of these trajectories changes when a ``fold" caustic is crossed. Classical physics clearly structures the form of the propagator. Note that the central figure includes two diagonal lines along which the amplitude is discontinuous. These are a result of the sharp cutoff in energy, and provide further motivation to work with smooth regulators as advocated in section \ref{smo}. Finally, note a qualitative difference between the exact propagator and its semiclassical approximation at the outside of the innermost fold caustic. Whereas the exact propagator includes a set of fringes, these are absent in the semiclassical approximation. These fringes are the consequence of a relevant complex saddle which exits the space of real paths at the fold and cusp caustics. Near the caustic, the complex saddle point makes a significant contribution. Quantum tunneling phenomena are another situation in which the contribution of relevant, complex saddle points cannot be ignored. Our method allows one to systematically improve these calculations, for any $\hbar$, by integrating over the eigenthimbles associated with relevant classical saddles. In a companion paper, we implement a numerical algorithm to establish the relevance or otherwise of complex saddles in quantum mechanical theories. We show there, in particular, that the complex saddles discussed in the footnote below Eq.~(\ref{ex2}) are irrelevant in the Picard-Lefschetz sense.

For the real time path integral, all real, classical solutions which satisfy the boundary conditions are relevant, by the Picard-Lefschetz criterion. However, when we analytically continue to Euclidean time, $T\rightarrow -i T_E$, with $T_E$ real, these saddles typically become irrelevant. The classical solutions (\ref{ex2}) for the quartic oscillator, for example, satisfy $x(0)=x(1)=0$. They satisfy the equations of motion and the same boundary conditions for any complex $T$. So we can continuously rotate the time to Euclidean values via $T=|T|e^{-i\theta}$, with $0\leq \theta \leq {\pi\over 2}$, and these solutions remain legitimate saddles. To see whether they remain relevant, we calculate their height function $h={\rm Re}[i S_C]\propto {\rm Re}[i/T^3] \sim - \sin (3 \theta)$. For $0<\theta<{\pi\over 3}$, $h$ is negative and the saddles may be relevant. However, when $\theta$ exceeds ${\pi\over 3}$, the height function is positive and the saddles are definitely irrelevant. Hence, the intricate, non-perturbative interference phenomena exhibited in Fig.~\ref{fig:quantumint}, which are so economically captured by real classical saddles, would be exponentially difficult to reproduce in Euclidean calculations.

{\bf Acknowledgements:}  We thank Sam Bateman, Luigi Del Debbio, Einan Gardi, Tony Kennedy, Minhyong Kim, Kostas Tsanavaris, Roman Zwicky and Ziwei Wang and other colleagues at the Higgs Centre for Theoretical Physics and the International Centre for Mathematical Sciences at Edinburgh for encouraging comments and helpful suggestions. We specially thank Tudor Dimofte for suggesting we use Paley-Wiener theorems to define suitable smooth regulators. The work of JF and NT is supported by the STFC Consolidated Grant `Particle Physics at the Higgs Centre,' and, respectively, by a Higgs Fellowship and the Higgs Chair of Theoretical Physics at the University of Edinburgh. Research at Perimeter Institute is supported by the Government of Canada, through Innovation, Science and Economic Development, Canada and by the Province of Ontario through the Ministry of Research, Innovation and Science.  

\bibliographystyle{utphys}
\bibliography{Library}

\providecommand{\href}[2]{#2}\begingroup\raggedright\begin{thebibliography}{10}

\bibitem{Feynman:1948}
R.~P. Feynman, ``Space-time approach to non-relativistic quantum mechanics,''
{\em Rev. Mod. Phys.} {\bf 20} (1948)  367--387.
%%CITATION = RMPHA,20,367;%%.

\bibitem{Feynman:1965}
R.~P. Feynman and A.~R. Hibbs, {\em {Quantum mechanics and path integrals}}.
\newblock International series in pure and applied physics. McGraw-Hill, New
  York, NY, 1965.

\bibitem{Grosche:1998}
C.~Grosche and F.~Steiner, {\em {Handbook of Feynman Path Integrals}}.
\newblock Springer Tracts in Modern Physics. Springer, Berlin, 1998.

\bibitem{Klauder:2003}
J.~R. {Klauder}, \href{http://dx.doi.org/10.1142/9789812795106\_0005}{``{The
  Feynman Path Integral: An Historical Slice},''} in {\em A Garden of Quanta:
  Essays in Honor of Hiroshi Ezawa. Edited by Arafune, J. {\it et al}.},
  pp.~55--76.
\newblock World Scientific, 2003.

\bibitem{Klauder:2010}
J.~Klauder, {\em A Modern Approach to Functional Integration}.
\newblock Applied and Numerical Harmonic Analysis. Birkh{\"a}user, Boston, USA,
  2010.

\bibitem{Kac:1949}
M.~{Kac}, ``{On distributions of certain Wiener functionals},'' {\em
  Transactions of the American Mathematical Society} {\bf 65} (1949)  1--13.

\bibitem{Glimm:2012}
J.~Glimm and A.~Jaffe, {\em Quantum Physics: A Functional Integral Point of
  View}.
\newblock Springer New York, 2012.

\bibitem{Tannor:2007}
D.~J. {Tannor}, {\em {Introduction to Quantum Mechanics: A Time-Dependent
  Perspective}}.
\newblock University Science Books, USA, 2007.

\bibitem{Alexandru:2020wrj}
A.~Alexandru, G.~Basar, P.~F. Bedaque, and N.~C. Warrington, ``{Complex paths
  around the sign problem},''
  \href{http://dx.doi.org/10.1103/RevModPhys.94.015006}{{\em Rev. Mod. Phys.}
  {\bf 94} (2022) no.~1, 015006}, \href{http://arxiv.org/abs/2007.05436}{{\tt
  arXiv:2007.05436 [hep-lat]}}.

\bibitem{Mondaini:2021ywk}
R.~Mondaini, S.~Tarat, and R.~T. Scalettar, ``{Quantum Critical Points and the
  Sign Problem},'' \href{http://arxiv.org/abs/2108.08974}{{\tt arXiv:2108.08974
  [cond-mat.str-el]}}.

\bibitem{Teitelboim:1982}
C.~{Teitelboim}, \href{http://dx.doi.org/10.1103/PhysRevD.25.3159}{``{Quantum
  mechanics on the gravitational field},''{\em \prd} {\bf 25} (June, 1982)
  3159--3179}.

\bibitem{Teitelboim:1983}
C.~{Teitelboim},
  \href{http://dx.doi.org/10.1103/PhysRevD.28.297}{``{Proper-time gauge in the
  quantum theory of gravitation},''{\em \prd} {\bf 28} (July, 1983)  297--309}.

\bibitem{Wheeler:1987}
J.~A. {Wheeler}, ``{Superspace and the Nature of Quantum Geometrodynamics},''
  in {\em Quantum Cosmology}, L.~Z. {Fang} and R.~{Ruffini}, eds., vol.~3,
  p.~27.
\newblock 1987.

\bibitem{Feldbrugge:2017}
J.~{Feldbrugge}, J.-L. {Lehners}, and N.~{Turok},
  \href{http://dx.doi.org/10.1103/PhysRevLett.119.171301}{``{No Smooth
  Beginning for Spacetime},''{\em \prl} {\bf 119} (Oct., 2017)  171301},
  \href{http://arxiv.org/abs/1705.00192}{{\tt arXiv:1705.00192 [hep-th]}}.

\bibitem{Feldbrugge:2017b}
J.~{Feldbrugge}, J.-L. {Lehners}, and N.~{Turok},
  \href{http://dx.doi.org/10.1103/PhysRevD.95.103508}{``{Lorentzian quantum
  cosmology},''{\em \prd} {\bf 95} (May, 2017)  103508},
  \href{http://arxiv.org/abs/1703.02076}{{\tt arXiv:1703.02076 [hep-th]}}.

\bibitem{Tucci:2019}
A.~{Di Tucci}, J.~{Feldbrugge}, J.-L. {Lehners}, and N.~{Turok},
  \href{http://dx.doi.org/10.1103/PhysRevD.100.063517}{``{Quantum
  incompleteness of inflation},''{\em \prd} {\bf 100} (Sept., 2019)  063517},
  \href{http://arxiv.org/abs/1906.09007}{{\tt arXiv:1906.09007 [hep-th]}}.

\bibitem{Turok:2022}
N.~{Turok} and L.~{Boyle}, ``{Gravitational entropy and the flatness,
  homogeneity and isotropy puzzles},''{\em arXiv e-prints} (Jan., 2022)
  arXiv:2201.07279, \href{http://arxiv.org/abs/2201.07279}{{\tt
  arXiv:2201.07279 [hep-th]}}.

\bibitem{FT2}
J.~Feldbrugge and N.~{Turok}, ``{Computing real-time path integrals},'' {\em in
  preparation}  .

\bibitem{Bjorken:100769}
J.~D. Bjorken and S.~D. Drell, {\em {Relativistic quantum mechanics}}, ch.~6,
  pp.~78--89.
\newblock International series in pure and applied physics.
\newblock McGraw-Hill, New York, NY, 1964.

\bibitem{Kent:2013}
A.~{Kent}, ``{Path Integrals and Reality},''{\em arXiv e-prints} (May, 2013)
  arXiv:1305.6565, \href{http://arxiv.org/abs/1305.6565}{{\tt arXiv:1305.6565
  [quant-ph]}}.

\bibitem{Donadi:2021yhd}
S.~Donadi and S.~Hossenfelder, ``{A path integral over Hilbert space for
  quantum mechanics},'' \href{http://dx.doi.org/10.1016/j.aop.2022.168827}{{\em
  Annals Phys.} {\bf 440} (2022)  168827},
  \href{http://arxiv.org/abs/2110.07168}{{\tt arXiv:2110.07168 [quant-ph]}}.

\bibitem{Witten:2010}
E.~{Witten}, ``{A New Look at the Path Integral of Quantum Mechanics},'' {\em
  arXiv e-prints} (2010)  , \href{http://arxiv.org/abs/1009.6032}{{\tt
  arXiv:1009.6032 [hep-th]}}.

\bibitem{Basar:2013}
G.~Basar, G.~V. Dunne, and M.~Unsal, ``{Resurgence theory, ghost-instantons,
  and analytic continuation of path integrals},'' {\em JHEP} {\bf 10} (2013)
  041, \href{http://arxiv.org/abs/1009.6032}{{\tt arXiv:1009.6032 [hep-th]}}.

\bibitem{Vassiliev}
V.~A. Vassiliev, {\em {Applied Picard-Lefschetz Theory}}, vol.~97 of {\em
  Mathematical Surveys and Monographs}.
\newblock American Mathematical Society, USA, 2002.

\bibitem{Keller}
J.~B. Keller, ``{Corrected Bohr-Sommerfeld quantisation conditions for non
  separable systems},'' {\em Annals of Physics} {\bf 4} (1958)  180--188.

\bibitem{Maslow}
V.~P. Maslow and M.~V. Fedoriuk, {\em {Semiclassical approximation in quantum
  mechanics}}, vol.~7 of {\em Mathematical physics and applied mathematics}.
\newblock D. Reidel, Dordrecht, Holland, 1981.

\bibitem{Dunne:2016nmc}
G.~V. Dunne and M.~\"Unsal, ``{New Nonperturbative Methods in Quantum Field
  Theory: From Large-N Orbifold Equivalence to Bions and Resurgence},''
  \href{http://dx.doi.org/10.1146/annurev-nucl-102115-044755}{{\em Ann. Rev.
  Nucl. Part. Sci.} {\bf 66} (2016)  245--272},
  \href{http://arxiv.org/abs/1601.03414}{{\tt arXiv:1601.03414 [hep-th]}}.

\bibitem{Costin:2020}
O.~Costin and G.~V. Dunne,
  \href{http://dx.doi.org/10.1016/j.physletb.2020.135627}{``{Physical resurgent
  extrapolation},''{\em Physics Letters B} {\bf 808} (Sept., 2020)  135627},
  \href{http://arxiv.org/abs/2003.07451}{{\tt arXiv:2003.07451 [hep-th]}}.

\bibitem{Bajnok:2021dri}
Z.~Bajnok, J.~Balog, and I.~Vona, ``{Analytic resurgence in the O(4) model},''
  \href{http://dx.doi.org/10.1007/JHEP04(2022)043}{{\em JHEP} {\bf 04} (2022)
  043}, \href{http://arxiv.org/abs/2111.15390}{{\tt arXiv:2111.15390
  [hep-th]}}.

\bibitem{Pazarbasi:2021ifb}
C.~Pazarba\c{s}\i{} and M.~\"Unsal, ``{Cluster Expansion and Resurgence in the
  Polyakov Model},''
  \href{http://dx.doi.org/10.1103/PhysRevLett.128.151601}{{\em Phys. Rev.
  Lett.} {\bf 128} (2022) no.~15, 151601},
  \href{http://arxiv.org/abs/2110.05612}{{\tt arXiv:2110.05612 [hep-th]}}.

\bibitem{Berry:1990}
M.~V. {Berry} and C.~J. {Howls},
  \href{http://dx.doi.org/10.1098/rspa.1990.0111}{``{Hyperasymptotics},''{\em
  Proceedings of the Royal Society of London Series A} {\bf 430} (Sept., 1990)
  653--668}.

\bibitem{Berry:1991}
M.~V. {Berry} and C.~J. {Howls},
  \href{http://dx.doi.org/10.1098/rspa.1991.0119}{``{Hyperasymptotics for
  Integrals with Saddles},''{\em Proceedings of the Royal Society of London
  Series A} {\bf 434} (Sept., 1991)  657--675}.

\bibitem{Berry:1993}
M.~V. {Berry} and C.~J. {Howls},
  \href{http://dx.doi.org/10.1098/rspa.1993.0134}{``{Unfolding the High Orders
  of Asymptotic Expansions with Coalescing Saddles: Singularity Theory,
  Crossover and Duality},''{\em Proceedings of the Royal Society of London
  Series A} {\bf 443} (Oct., 1993)  107--126}.

\bibitem{Howls:1997}
C.~J. {Howls},
  \href{http://dx.doi.org/10.1098/rspa.1997.0122}{``{Hyperasymptotics for
  Multidimensional Integrals, Exact Remainder Terms and the Global Connection
  Problem},''{\em Proceedings of the Royal Society of London Series A} {\bf
  453} (Nov., 1997)  2271--2294}.

\bibitem{Strichartz}
R.~S. Strichartz, {\em {A guide to distribution theory and Fourier
  transforms}}, ch.~7, pp.~119--120.
\newblock I.
\newblock World Scientific, Singapore, 1994.

\bibitem{MorseandFeshbach}
P.~M. Morse and H.~Feshbach, {\em {Methods of Theoretical Physics}}, vol.~I,
  ch.~6, p.~760.
\newblock McGraw Hill, New York, 1953.

\bibitem{Levit:1977}
M.~Levit and U.~Smilansky, ``{A theorem on infinite products of eigenvalues of
  Sturm-Liouville type operators},'' {\em Proceedings of the American
  Mathematical Society} {\bf 65} (1977)  299--302.

\bibitem{Forman:1987}
R.~Forman, ``{Functional determinants and geometry},'' {\em Inventiones
  Mathematicae} {\bf 88} (1987)  447--494.

\bibitem{McKane:1995}
A.~J. McKane and M.~B. Tarlie, ``{Regularization of functional determinants
  using boundary perturbations},'' {\em J Phys A} {\bf 28} (1995)  6931--6942.

\bibitem{Feldbrugge:2019arXiv190904632F}
J.~{Feldbrugge}, U.-L. {Pen}, and N.~{Turok}, ``{Oscillatory path integrals for
  radio astronomy},''{\em arXiv e-prints} (Sept., 2019)  arXiv:1909.04632,
  \href{http://arxiv.org/abs/1909.04632}{{\tt arXiv:1909.04632 [astro-ph.HE]}}.

\bibitem{2020arXiv201003089F}
J.~{Feldbrugge}, ``{Multi-plane lensing in wave optics},''{\em arXiv e-prints}
  (Oct., 2020)  arXiv:2010.03089, \href{http://arxiv.org/abs/2010.03089}{{\tt
  arXiv:2010.03089 [astro-ph.CO]}}.

\bibitem{Abramowitz}
M.~Abramowitz and I.~A. Stegun, {\em {Handbook of Mathematical Functions}},
  ch.~16, pp.~569--585.
\newblock {Dover}, New York, NY, 1965.

\bibitem{Fulling:2003}
S.~A. Fulling and K.~S. Güntürk, ``{Exploring the propagator of a particle in
  a box},'' {\em American Journal of Physics} {\bf 71} (2003)  55--63.

\bibitem{Subramanyan:2021}
V.~Subramanyan, S.~H. Suraj, S.~Vishveshwara, and B.~Bradley, ``{Physics of the
  inverted harmonic oscillator: from the lowest Landau level to event
  horizons},'' {\em Annals of Physics} {\bf 435} (2021)  168470.

\bibitem{Carreau:1990}
M.~{Carreau}, E.~{Farhi}, S.~{Gutmann}, and P.~F. {Mende},
  \href{http://dx.doi.org/10.1016/0003-4916(90)90125-8}{``{The functional
  integral for quantum systems with Hamiltonians unbounded from below},''{\em
  Annals of Physics} {\bf 204} (Nov., 1990)  186--207}.

\bibitem{Wheeler:1959}
K.~W. {Ford}, D.~L. {Hill}, M.~{Wakano}, and J.~A. {Wheeler},
  \href{http://dx.doi.org/10.1016/0003-4916(59)90025-9}{``{Quantum effects near
  a barrier maximum},''{\em Annals of Physics} {\bf 7} (July, 1959)  239--258}.

\bibitem{Barton:1986}
G.~{Barton}, \href{http://dx.doi.org/10.1016/0003-4916(86)90142-9}{``{Quantum
  mechanics of the inverted oscillator potential},''{\em Annals of Physics}
  {\bf 166} (Feb., 1986)  322--363}.

\bibitem{1978mmcm.book.....A}
V.~I. {Arnold}, {\em {Mathematical methods of classical mechanics}}.
\newblock 1978.

\bibitem{2022arXiv220412004J}
D.~L. {Jow}, U.-L. {Pen}, and J.~{Feldbrugge}, ``{Regimes in astrophysical
  lensing: refractive optics, diffractive optics, and the Fresnel scale},''{\em
  arXiv e-prints} (Apr., 2022)  arXiv:2204.12004,
  \href{http://arxiv.org/abs/2204.12004}{{\tt arXiv:2204.12004 [astro-ph.HE]}}.

\bibitem{Bartle:1995}
R.~G. Bartle, {\em The Elements of Integration and Lebesgue Measure}.
\newblock John Wiley \& Sons, New York, 1995.

\bibitem{Hunt:1992}
B.~R. Hunt, T.~Sauer, and J.~A. Yorke, ``{Prevalence: a translation-invariant
  "almost every" on infinite-dimensional space},'' {\em Bull. Amer. Math. Soc.
  (N.S.)} {\bf 27} (2006) no.~2, 217--238.

\bibitem{Costello}
K.~Costello, {\em {Renormalization and Effective Field Theory}}, vol.~170 of
  {\em Mathematical Surveys and Monographs}, p.~5.
\newblock American Mathematical Society, 2011.

\bibitem{Simon}
B.~Simon, {\em {Functional Integration and Quantum Physics}}.
\newblock {Academic Press}, New York, 1979.

\bibitem{Madelung}
E.~Madelung, ``{Das elektrische Feld in Systemen von regelmäßig angeordneten
  Punktladungen},'' {\em Phys. Z.} {\bf XIX} (1918)  524--533.

\bibitem{Bailey}
D.~Bailey, J.~Borwein, V.~Kapoor, and E.~Weisstein, ``{Ten Problems in
  Experimental Mathematics},'' {\em The American Mathematical Monthly} {\bf
  113} (2006) no.~6, 481.

\end{thebibliography}\endgroup


\providecommand{\href}[2]{#2}\begingroup\raggedright\endgroup

%\begin{thebibliography}{99}

%\bibitem{Parnell}

%\end{thebibliography}

\appendix

%%%%%%%%%%%%%%%%%%%%%%%%%%%%%%%%%%%%%%%%%%%%%%%%%%%%%%%%%%%%%%%%%%%%%%%%%%%%%%%%%%%%
\section{Measure theory on the space of paths}
\label{measuretheory}
The Wiener measure and the Brownian bridge are famous measures on the space of paths which facilitate the construction of mathematically rigorous path integrals. These measures are central to the Feynman-Kac formula for the Euclidean path integral. We also use them in this paper to show that the real time (or Lorentzian) path integral exists, {\it i.e.}, it defines a complex measure. In this appendix, we briefly summarize the key concepts in measure and integration theory as applied to the space of paths.

\begin{figure}
\centering
\includegraphics[width=0.9\textwidth]{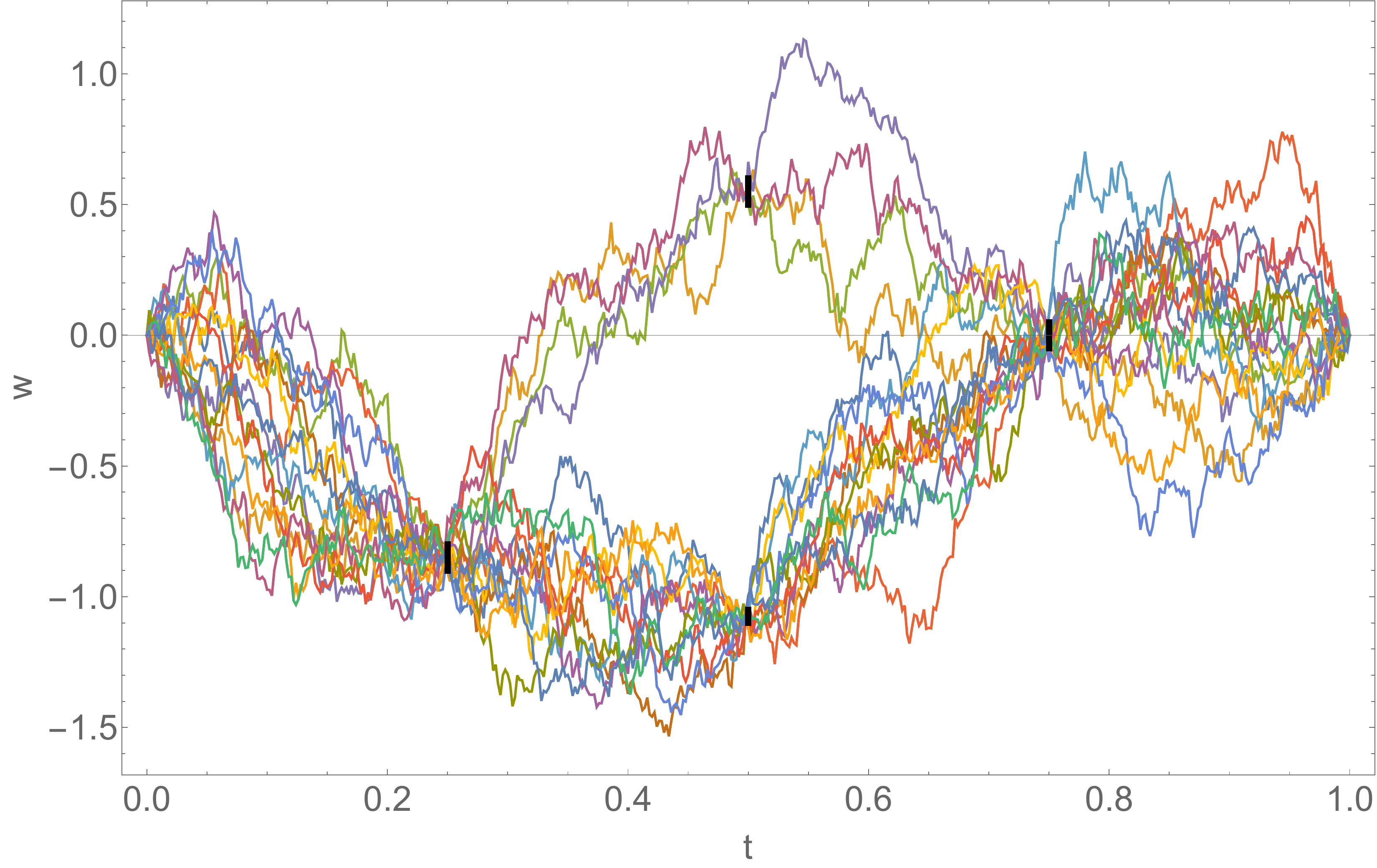}
\caption{Realizations of a Brownian bridge process passing through a series of slits.}\label{fig:slits}
\end{figure}

Measure theory starts with the theory of subsets and in particular $\sigma$-algebras (see, {\it e.g.},~\cite{Bartle:1995} for an introduction). Consider the space $\Omega$ of all paths $w:[0,1]\to \mathbb{R}$ starting at $w(0)=0$ and ending at $w(1)=0$, {\it i.e.} ``going nowhere." In the space $\Omega$, we define the subset of paths
\begin{align}
Q=\{w \in \Omega\,|\, a < w(t') < b\}
\end{align}
which pass through the interval (or ``slit") $(a,b)$ at some time $0< t' <1$. $\Omega$ is itself such a subset,  defined at any particular time $t'$ with the interval comprising the entire real line. By taking finite unions and intersections of such subsets, we generate a set of subsets of $\Omega$. In particular, we generate a subset consisting of the paths which pass through  a set of slits, $w(t_i) \in (a_i,b_i)$, $i=1,\dots n$, defined at any ordered set of intermediate times $0< t_i <1$,
\begin{align}
Q=\{w \in \Omega\,|\, a_i < w(t_i) < b_i, 0 < t_1 < \dots < t_n<1\}\,.
\label{q2}
\end{align}
See fig.\ \ref{fig:slits} for an illustration. By extending this construction to all countable unions and intersections, including infinite sets of subsets, we generate the Borel field $\mathcal{A}$ of subsets of $\Omega$. This Borel field $\mathcal{A}$ is known as a $\sigma$-algebra, as it (i) includes the total space $\Omega \in \mathcal{A}$, (ii) is closed under the complement operation, \textit{i.e.}, when $A \in \mathcal{A}$ then also $A^c = \Omega \backslash A \in \mathcal{A}$, and (iii) is closed under countable unions, \textit{i.e.}, when $A_n \in \mathcal{A}$ for $n\in \mathbb{N}$, then also $\cup_{n=1}^\infty A_n \in \mathcal{A}$. 

A measure on a $\sigma$-algebra is a map which assigns a `size' to any subset in the $\sigma$-algebra. Formally, a measure $\mu$ on the $\sigma$-algebra $\mathcal{A}$ is a mapping $\mu: \mathcal{A} \to [0,\infty]$ for which (i) the measure of the empty set vanishes $\mu(\emptyset)=0$, and (ii) the measure of a union of pairwise disjoint sets coincides with sum of the measure of the sets, \textit{i.e.}, for pairwise disjoint $A_n \in \mathcal{A}$ for $n\in \mathbb{N}$, we require $\mu(\cup_{n=1}^\infty A_n)=\sum_{n=1}^\infty \mu(A_n)$. The second condition, known as the \textit{countable additivity}, is of central importance to the definition of the integral over $\Omega$ as we will see below.

In finite dimension $D$, the product of Lebesgue measures 
\begin{align}
\mu_D(A)=\int_A \prod_{i=1}^D \mathrm{d}w_i
\end{align}
on the subsets $ A \subset \mathbb{R}^D$ is a Lebesgue measure. It is clearly translation invariant. However, a well-known theorem shows that for the kinds of space we are interested in (technically, separable Banach spaces), in infinite dimensions every translation-invariant Lebesgue measure which is not identically zero has the property that every open set has infinite measure. Such a measure would be useless for physics. 

The proof is elementary~\cite{Hunt:1992}. Suppose that an open ball of radius $\epsilon$ has finite measure. Using elementary geometry, one sees that, given an unlimited number of dimensions, an infinite number of open balls of radius $\epsilon/3$ can be fitted inside the original ball (the number $3$ is unimportant: all that matters is that it is finite). By translation invariance, every $\epsilon/3$-ball has identical measure. But the sum of their measures, being bounded by that of the $\epsilon$-ball, is finite. Hence every $\epsilon/3$-ball must have zero measure. However, the whole space is separable so it can be covered with a countable collection of $\epsilon/3$-balls. Thus the whole space has measure zero. The only way out of this conclusion is to suppose that the measure of the $\epsilon$-ball is infinite. 

Although this no-go theorem is well-known to mathematicians, it is seldom mentioned in theoretical physics textbooks. One exception is a recent mathematical textbook on renormalization and effective field theory, which states ``The non-existence of a [translation-invariant] Lebesgue measure on an infinite-dimensional vector space is one of the fundamental difficulties of quantum field theory"~\cite{Costello}. When the author subsequently refers to the na\"{\i}ve path integral measure (as he frequently does) he usually calls it ``the non-existent Lebesgue measure"!

The Wiener and Brownian bridge measures circumvent the no-go theorem by dropping the assumption of translational invariance. The Wiener process is the scale-invariant limit of a random walk starting at $w(0)=0$, where the step size tends to zero while the number of steps tends to infinity. The Brownian bridge, which we use in this paper, is a similar process where not only the starting point but also the endpoint is pinned to zero, \textit{i.e.}, $w(1)=0$. For the Brownian bridge, the general subset $Q$ as in (\ref{q2}) is assigned the probability
\begin{align}
\mu_B(Q) =& \left({\sqrt{2 \pi} W\over \prod_{i=1}^{N+1}\left( W \sqrt{2 \pi (t_{i}-t_{i-1})}\right)}\right)
\int_{a_1}^{b_1} \dots \int_{a_{N}}^{b_{N}}e^{-{\displaystyle \sum_{i=1}^{N+1} }\frac{(w_i-w_{i-1})^2}{2W^2(t_i-t_{i-1})}}\mathrm{d}w_1 \dots \mathrm{d}w_{N}\,,
\label{m1}
\end{align}  
where $w_i \equiv w(t_i)$, with $w_0=w_{N+1}=0$, $t_0=0$, $t_{N+1}=1$ and $0<t_1<t_2\dots<t_N<1$. In anticipation of the connection with quantum mechanics, we include a weight $W$ which parameterizes the strength of the ``jitter" in the walk. The measure (\ref{m1}) extends to the Brownian bridge measure on the Borel $\sigma$-algebra $\mathcal{A}$ of the space of paths $\Omega$. (Note that there exist several different but equivalent definitions including one based on the Bochner-Minlos theorem using the characteristic functionals of the Brownian bridge process~\cite{Simon}). As the measure of the complete set of paths is unity, \textit{i.e.}, $\mu_B(\Omega)=1$, the Brownian bridge measure is a probability measure. The measure $\mu_B(Q)$ is then simply the probability for the Brownian bridge process to pass through the intervals which define $Q$. From its construction it follows that the space for which the measure has compact support consists of the almost everywhere continuous and almost nowhere differentiable paths. That is to say, with probability one a path is continuous but not differentiable, at every intermediate time. Finally, note that if we take equally spaced times, $t_{i+1}-t_i=\Delta t$, the measure (\ref{m1}) is proportional to the time-discretized Euclidean path integral for a free particle. This suggests an intimate link between the Brownian bridge and Feynman's path integral.

Given the countably additive Brownian bridge measure $\mu_B$, we construct the corresponding integral as follows. Let $f$ be the \textit{simple} function defined as
\begin{align}
f = \sum_{i=1}^r \alpha_i \bm{1}_{A_i}
\end{align}
for some positive coefficients $\alpha_1,\dots,\alpha_r \geq 0$ and the pairwise disjoint sets $A_1, \dots,A_r \in \mathcal{A}$. The identity function $\bm{1}_A(x)$ is unity when $x \in A$ and vanishes otherwise. The integral over this simple function is defined as the sum of the product of the coefficients and the measures of the corresponding sets,
\begin{align}
\int_{\Omega} f\, \mathrm{d}\mu_B = \sum_{i=1}^r \alpha_i \mu_B(A_k).
\end{align}
Given a more general non-negative function $f$ on the space of paths $\Omega$, we define the integral as the supremum over the integral of the simple functions it dominates
\begin{align}
\int_{\Omega} f\, \mathrm{d}\mu_B = \text{sup}\left\{ \int_{\Omega} g\, \mathrm{d}\mu_B\, \bigg|\, \text{ where } g \text{ is simple and } 0 \leq g \leq f\right\}\,.
\end{align} 
In general, any non-negative function of this type can be approached as a limit of ever increasing simple functions. We define the integral over a general function on the space of paths as the difference of its positive and negative parts, \textit{i.e.}, when we split the function in its positive and negative parts $f=f_+ - f_i$ with $f_+,f_- \geq 0$, we define the integral as
\begin{align}
\int_\Omega f\mathrm{d}\mu_B =\int_\Omega f_+\mathrm{d}\mu_B - \int_\Omega f_-\mathrm{d}\mu_B\,,
\end{align}
assuming both $\int f_+\mathrm{d}\mu_B$ and $\int f_-\mathrm{d}\mu_B$ are finite. It is more or less clear from these definitions that given a measure, such as the Brownian bridge or Wiener measures, one can construct a more general measure by simply multiplying the measure with a positive, integrable function. In the main text, this is exactly what we do in equation (\ref{g9}) when we multiply the Brownian bridge measure by the function $g$ which, by definition is positive and which we later show is bounded above by unity. Having shown that this positive measure exists, we assume it can be straightforwardly extended to a complex measure which it bounds. 

The definition (\ref{m1}) of the Brownian bridge measure leads to a practical method for evaluating the integral over a functional $f$ which only depends on the random process at a series of times $0 =t_0  < t_1 < \dots < t_{N+1}=1$. In this case, the infinite-dimensional integral reduces to the $N$-dimensional integral
\begin{align}
\int f \mathrm{d}\mu_B = & \left({\sqrt{2 \pi} W\over \prod_{i=1}^{N+1}\left( W \sqrt{2 \pi (t_{i}-t_{i-1})}\right)}\right)
\int  f(w_1,\dots,w_{N})e^{-{\displaystyle \sum_{i=1}^{N+1} }\frac{(w_i-w_{i-1})^2}{2 W^2 (t_i-t_{i-1})}}\mathrm{d}w_1 \dots \mathrm{d}w_{N}.
\end{align}

The Brownian bridge is a Gaussian process with mean zero. It follows that all of its properties are completely determined by the covariance matrix 
\begin{align}
\langle w(t') w(t) \rangle_B &=W^2 \, t(1-t') \qquad 0<t<t'<1,\cr
&=W^2 \,t'(1-t) \qquad 0<t'<t<1.
\label{2pt}
\end{align}
This covariance is reproduced by taking $w(t)$ to be a linear superposition of sinusoidal waves,  $w(t) = \sum_{k=1}^\infty w_k \sqrt{2} \sin(k\pi t)$, where the $w_k$ are independent, normally distributed random variables with vanishing mean and variance $W^2/(k \pi)^2$. (The Karhunen-Lo\`eve theorem indicates this choice of basis functions to be optimal). The corresponding statistical covariance is $\displaystyle  \langle w(t') w(t) \rangle = W^2  \sum_{k=1}^\infty 2/(k \pi)^2 \sin (k \pi t')\sin (k \pi t)$, which is just (\ref{2pt}) expressed as a Fourier series.  In this representation, the Brownian bridge measure takes the form
\begin{align}
\mu_B(A) \equiv \int_A \, \prod_{k=1}^\infty \frac{e^{- w_k^2/(2\sigma_k^2)}}{\sqrt{2\pi}\sigma_k} \,\mathrm{d}w_k; \qquad \sigma_k=W/(k \pi).
\label{ea5}
\end{align}
Note that this expression is only formal as the sum appearing in the exponent will typically diverge for Brownian bridges (because typical paths are non-differentiable) and the infinite product $\prod_{k=1}^\infty \mathrm{d}w_k/(\sqrt{2 \pi} \sigma_k)$ which multiplies it  is not a meaningful measure, as explained above. The integral expression only makes sense as a whole because the two effects are finely balanced. Nevertheless, this explicit representation is extremely useful.

From (\ref{2pt}) it follows that $\displaystyle \lim_{\epsilon \to 0} \langle (w(t+\epsilon)-w(t))^2 \rangle_B= \epsilon$, indicating that typical paths contributing to the measure are continuous. However, by differentiating  (\ref{2pt}) one finds the velocity-velocity correlator $\langle \dot{w}(t') \dot{w}(t) \rangle_B=W^2 (\delta(t'-t)-1)$. In this case, $\displaystyle \lim_{\epsilon \to 0} \langle (\dot{w}(t+\epsilon)-\dot{w}(t))^2 \rangle_B=2 W^2\delta(0)$ so typical paths have discontinuous velocities. 

In the main text of the paper, before equation (\ref{g1w}), we explained how the free particle path integral produces an expression just like (\ref{ea5}), when we set $x(t)=e^{i \pi/4} w(t)$, rotating the real path to the Wiener thimble and expressing it as a Fourier series.  The precise relation between the measure obtained from the free particle path integral and (\ref{ea5}) is simply
\begin{align}
W = \sqrt{\hbar T\over M},
\label{ea6}
\end{align}
which is clearly a measure of the ``quantum spreading" in the process.  We infer that 
\begin{align}
\langle x(t') x(t) \rangle &=i {\hbar T\over M} \, t(1-t') \qquad 0<t<t'<1,\cr
&=i {\hbar T\over M} \,t'(1-t) \qquad 0<t'<t<1.
\label{ea7}
\end{align}
The appearance of $i$ times Planck's constant suggests a connection with Heisenberg's quantum commutator of the position and momentum, and the uncertainty relation. This is indeed the case, as we shall now see. In fact, we'll see that implementing the correct uncertainty relation with a statistical ensemble actually {\it requires} an infinite-dimensional measure. 

In quantum mechanics, the time-dependent Heisenberg operators $\hat{x}(t')$ and $\hat{x}(t)$, taken at unequal times $t'$ and $t$, do not commute. So if we are to identify the correlator (\ref{ea7}) with a quantum correlator, the latter must involve a particular ordering. In the Heisenberg picture, the state vectors are time-independent. The ket vector vector $|0,0\rangle$ (the first and second arguments indicating $x$ and $t$, respectively) is annihilated by the position operator at time $t=0$: $\hat{x}(0)|0,0\rangle=0$. Likewise, the bra vector $\langle0,1|$ is annihilated by the position operator at $t=1$:  $\langle0,1| \hat{x}(1)=0$. It is therefore natural to attempt to relate the statistical correlator in (\ref{ea7}) with the {\it time-ordered} quantum correlator
\begin{align}
G_{xx}(t',t)\equiv {\langle 0,1| T\left(\hat{x}(t') \hat{x}(t)\right)|0,0 \rangle\over \langle 0,1| 0,0 \rangle} &\equiv{ \langle 0,1| \hat{x}(t') \hat{x}(t)|0,0 \rangle\over \langle 0,1| 0,0 \rangle}, \qquad t'>t,\cr
&\equiv {\langle 0,1| \hat{x}(t) \hat{x}(t')|0,0 \rangle\over \langle 0,1| 0,0 \rangle}, \qquad t>t',
\label{ea8}
\end{align}
which, like (\ref{ea7}), vanishes both at $t'=0,1$ and at $t=0,1$. We normalize it by dividing by the state overlap with no operator inserted. The Heisenberg operator $\hat{x}(t)$ obeys the free particle equation of motion ${\mathrm{d}^2\hat{x}/ \mathrm{d}t^2}=0$, so $G_{xx}(t',t)$ must be linear in both $t'$ and $t$ except at $t'=t$ where the time ordering comes into play. To understand what happens there, calculate the change in $t'$ derivative, across $t'=t$:  
\begin{align}
\left[\partial_{t'} G_{xx}(t',t)\right]_{t'=t_-}^{t'=t_+} &={\langle 0,1| \dot{\hat{x}}(t) \hat{x}(t)- \hat{x}(t)\dot{\hat{x}}(t) |0,0 \rangle\over \langle 0,1| 0,0 \rangle} = {T\over M} {\langle 0,1|\left[\hat{p}(t),\hat{x}(t)\right]|0,0 \rangle\over \langle 0,1| 0,0 \rangle} = -i \hbar {T\over M},
\label{ea9}
\end{align}
where the velocity is related to the momentum via $\hat{p}= M \dot{\hat{x}}/T$. Since $G_{xx}(t',t)$ obeys the same equations of motion, boundary conditions and jump condition as the analytically continued statistical correlator, we conclude that $ G_{xx}(t',t)= \langle x(t') x(t) \rangle$, given in (\ref{ea7}).

Notice that the equal time quantum correlator $G_{xx}(t,t)$ is in fact imaginary for all $0<t<1$. This might seem surprising since the operator $\hat{x}^2(t)$ is Hermitian. However, there is no contradiction because the states $|0,0 \rangle$ and  $\langle 0,1|$ are {\it not} Hermitian conjugates. Notice also that Heisenberg's commutation relation between $\hat{x}(t)$ and $\hat{p}(t)$ arises as a direct consequence of the discontinuity in the $t'$ derivative of the statistical correlator $\langle x(t') x(t) \rangle$ at $t'=t$, which requires an  infinite number of Fourier modes to represent it. By differentiating  $G_{xx}(t',t)$ with respect to  $t$ and $t'$, we find the  quantum time-ordered momentum-momentum correlator $G_{pp}(t',t)=i \hbar {M\over T}(\delta(t'-t)-1)$ which is also imaginary. It diverges at equal times as  a symptom of the high frequency ``jitter" on typical contributing paths. It is an instructive exercise to derive these results for $G_{xx}(t',t)$ and $G_{pp}(t',t)$ in the Schr\"odinger picture, using the free-particle propagator $K(x_1,t_1-t_0;x_0)=e^{i M(x_1-x_0)^2/\left(2 \hbar (t_1-t_0)\right)}\sqrt{M/(2 \pi i \hbar (t_1-t_0))} $ and performing integrals over $x$ via the contour rotation $x=e^{i \pi/4} w$.  

The Brownian bridge measure (\ref{ea5}) is also the basis for the definition of the Euclidean path integral for quantum mechanics, {\it i.e.} the Feynman-Kac formula
\begin{align}
\frac{\int e^{-(1/\hbar)\int (M\dot{x}^2/2+V(x))\mathrm{d} \tau } \mathcal{D}x}{\int e^{-(1/\hbar)\int (M\dot{x}^2/2) \mathrm{d}\tau } \mathcal{D}x} \equiv \int  e^{-(1/\hbar)\int V(x)\mathrm{d}\tau }\mathrm{d}\mu_B(x)
\end{align}
where $\tau$ is the Euclidean time and dots are $\tau$ derivatives. This formulation is extremely useful, for example, for calculating stationary states and energy spectra when the potential $V(x)$ is bounded from below. ``Vacuum" boundary conditions are usually imposed, namely $x(\tau)$ is assumed to tend to the global minimum of the potential as $\tau\rightarrow \pm\infty$. Taking this vacuum to be located at $x=0$, for example, $x$ is the fluctuation and is integrated over in the corresponding Brownian bridge measure. (Setting $\tau= i T t$, with $T$ and $t$ as used elsewhere in this paper, $x(t)$ is identified with $w(t)$ (with no complex rotation) and $T$ is taken to be negative imaginary). We thus formally identify $\mathrm{d}\mu_B \propto   e^{-(1/\hbar) \int (\dot{x}^2/2)\mathrm{d}\tau}\mathcal{D}x$. The smoothing due to the kinetic term and wildness of the infinite product ``measure" are beautifully balanced in the Brownian bridge.

%%%%%%%%%%%%%%%%%%%%%%%%%%%%%%%%%%%%%%%%%%%%%%%%%%%%%%%%%%%%%%%%%%%%%%%%%%%%%%%%%%%%
\section{Fubini's theorem and the dominated convergence theorem}
\label{theorems}
In this appendix, we illustrate and state two key theorems which we make use of in this paper. These theorems apply to absolutely convergent integrals, or equivalently to integrable functions. For a detailed mathematical treatment see \cite{Bartle:1995}.

Multidimensional integrals are usually defined without specifying the order in which partial integrals are to be performed. For such a definition to make sense, the result must be independent of the order in which the partial integrals are taken. 

Similar issues arise in the more elementary context of alternating series. Consider, for example, the geometric series
\begin{align}
S_N = \sum_{n=1}^N \frac{(-1)^{n+1}}{2^n}
\end{align}
which converges to $1/3$ as $N\to \infty$. This result is independent of the order in which the terms are combined because the series converges absolutely as $\sum_{n=1}^\infty 2^{-n}=1$. On the other hand, the series
\begin{align}
S_N = \sum_{n=1}^N \frac{(-1)^{n+1}}{n}
\end{align}
converges, by the alternating series test, to the natural logarithm $\log 2$ as $N\to \infty$. However, in this case the result {\it does} depend on the order in which the terms are included in the sum, because the absolute sum $\sum_{n =1}^\infty n^{-1}$ diverges. Just as in our discussion of the convergence of the discretized path integral for a free particle, in Section \ref{smo}, one might attempt to define such sums by introducing a cutoff which is then sent to infinity. As we found there, this procedure works well in one dimension. However, in higher dimensions, the cutoff method again fails badly. Such higher dimensional sums occur in calculations of the electrostatic binding energies of a charged ion in a crystal -- for example, a sodium or chlorine ion in a salt crystal -- where one has to perform a three-dimensional lattice sum. The physical binding energy of an ion is proportional to an infinite lattice sum known as the Madelung constant~\cite{Madelung}. To evaluate the sum, a naive approach would be to include only the charges contained in some large bounding surface. However, if the surface is not chosen carefully, the net charge scales as the square root of the area, leading to to an answer which oscillates with fixed amplitude as the radius of the surface is taken to infinity. For real salt crystals, any net charge on a finite crystal is rapidly neutralized by ions from the surrounding environment, hence the sum is regulated physically. There are also nice mathematical regularizations which yield a unique, regulator-independent result: for a recent discussion, see {\it e.g.}, \cite{Bailey}. 

Fubini's theorem assures us that the result of a higher-dimensional integral is independent of the order in which partial integrals are performed:\hfill\\
\noindent
\textbf{Fubini's theorem:} \textit{when a two-dimensional integral $\int f(x,y) \mathrm{d}(x,y)$ is absolutely convergent, \textit{i.e.}, $\int |f(x,y)| \mathrm{d}(x, y) <\infty$, the integral can be evaluated sequentially and is independent of the order in which partial integrals are taken, \textit{i.e.},
\begin{align}
\int f(x,y) \mathrm{d}(x,y) 
=
\int \left[\int f(x,y) \mathrm{d}x\right] \mathrm{d}y
=
\int \left[\int f(x,y) \mathrm{d}y\right] \mathrm{d}x\,.
\end{align}}

The necessity for absolute convergence is seen from the following counter-example. Consider the two-dimensional integral
\begin{align}
I=\int_{[0,1]^2} \frac{x^2-y^2}{(x^2+y^2)^2}\mathrm{d}(x,y),
\end{align}
which is absolutely divergent, \textit{i.e.}, $\int_{[0,1]^2} \left|\frac{x^2-y^2}{(x^2+y^2)^2}\right|\mathrm{d}(x,y)=\infty$. Integrating first with respect to $x$ and then $y$ yields the result $-\pi/4$ whereas integrating in the reverse order yields the opposite result $\pi/4$. Alternatively, using polar coordinates we obtain the representation $I= \iint (\cos(2\theta)/r )\mathrm{d}( r,\theta)$. Integrating first with respect to $\theta$ (with limits depending on $r$)  gives a vanishing answer as $\cos(2\theta)$ is odd around $\theta =\pi/4$.

The dominated convergence theorem is central to study of limits in integration theory:\hfill\\
\noindent
\textbf{Dominated convergence theorem:} \textit{let $(f_n)$ be a sequence of absolutely convergent functions which converges pointwise to the function $f$ as $n\to \infty$ and is dominated by an integrable function $g$, \textit{i.e.}, $|f_n(x)| < g(x)$ for all $x$ and $n$, with $\int |g(x)|\mathrm{d}x < \infty$, than $f$ is integrable (integrates absolutely) and 
\begin{align}
\lim_{n\to \infty} \int f_n(x)\mathrm{d}x = \int f(x)\mathrm{d}x\,.
\end{align}}

The theorem generalizes to continuous limits over continuous functions
\begin{align}
\lim_{y\to \infty }\int f(x,y)\mathrm{d}x= \int \lim_{y\to \infty} f(x,y)\mathrm{d}x
\end{align}
with an integrable function $g(x)$ dominating $|f(x,y)|$ for all $y$. This can be seen by considering a sequence $(y_n)$ which diverges to $\infty$ as $n\to \infty$. Using the dominated convergence theorem, we find
\begin{align}
\lim_{y\to \infty }\int f(x,y)\mathrm{d}x
= \lim_{n\to\infty} \int f(x,y_n)\mathrm{d}x
=\int \lim_{n\to\infty}  f(x,y_n)\mathrm{d}x
=\int \lim_{y\to\infty}  f(x,y)\mathrm{d}x\,.
\end{align}

The absolute convergence assumption is necessary, as can be seen with the following counter-example. Consider the sequence of functions $f_n(x)=n$ when $x \in (0,1/n)$ and $0$ otherwise, which converges pointwise to the zero function as $n\to \infty$. The limit does not commute with the integral as
\begin{align}
\lim_{n\to \infty} \int f_n(x)\, \mathrm{d}x = \lim_{n\to \infty} \int_{0}^{1/n}n\, \mathrm{d}x = \lim_{n\to \infty} 1 = 1\,,
\end{align}
while 
\begin{align}
\int \lim_{n\to \infty}  f_n(x)\, \mathrm{d}x =\int 0 \, \mathrm{d}x=0\,.
\end{align}
Note that this example is not in tension with the dominated convergence theorem because the envelope $g=1/x$ of the sequence $f_n$ is not integrable, \textit{i.e.}, the integral of the envelope diverges $\int g(x)\,\mathrm{d}x = \infty$.

\section{LFA descent thimbles and eigenthimbles for quartic oscillator}
\label{etapp}

In this Appendix, we give analytic expressions for the thimbles corresponding to the real saddles in the quartic oscillator, for the amplitude to ``go nowhere," in the lowest frequency approximation (or LFA).  The model is defined by the action $S=\int_0^1{1\over 2}(\dot{x}^2-x^4)\mathrm{d}t$ where we have, for simplicity, set the mass and the time $T$ to unity. Due to the scaling symmetry of the model, all of the real solutions as well as their steepest descent thimbles and eigenthimbles can be obtained by scaling from the solutions with $n=1$. In the LFA, we set $x(t)=x_1 \sin (\pi t)$ and ignore higher modes. To simplify formulae, we suppress the subscript $1$ in the following.  The action reads $S_{LFA}={1\over 4}(\pi^2 x^2 -{3\over 4} x^4)$. In addition to the trivial saddle at $x=0$, the action has two real saddles at $x=\pm \sqrt{2\over 3} \pi$. 

Writing $iS =h+i H$, the imaginary part $H$ is constant along a steepest descent trajectory. In one complex dimension, the constancy of $H$ uniquely defines the trajectory: by passing to polar coordinates, we can find it explicitly. For the trivial saddle $x=0$, the two halves of the steepest descent thimble are given by
\begin{align}
x(\theta)=\pm{2 \pi \over\sqrt{3}}  e^{i \theta} \sqrt{\cos 2\theta\over \cos 4 \theta}, \quad {\pi\over 4} <\theta<{3\pi \over 8},
\label{sd1}
\end{align}
with the saddle located at $\theta=\pi/4$ whereas for the nontrivial saddles $x=\pm \sqrt{2\over 3} \pi$, the steepest descent thimbles are given by
\begin{align}
x(\theta)= \pm {\sqrt[4]{2} \pi \over \sqrt{3}} e^{i \pi/8} (\cosh \theta+i \sinh \theta), \quad -\infty <\theta< \infty
\label{sd2}
\end{align}
with the saddles located at $\theta=-\tanh^{-1}(\tan(\pi/8))$. 

We can also find the eigenthimbles analytically, in the LFA approximation. For the trivial saddle, the two halves of the eigenthimble are given by 
\begin{align}
x(\theta)=\pm \sqrt{2\over 3} \pi e^{i \theta} \sqrt{\cos 2\theta\over \cos 4 \theta}, \quad {\pi\over 4} < \theta<{3\pi \over 8},
\label{et3}
\end{align}
with the saddle again located at $\theta=\pi/4$, whereas for the nontrivial saddles at $x= \pm \sqrt{2\over 3} \pi$, the eigenthimbles are 
\begin{align}
x(\theta)=\pm \left(\sqrt{2\over 3} \pi +{\pi e^{i(\theta-{\pi\over 4})}\over 2\sqrt{3} \cos 4 \theta} \left(3\sqrt{2}\sin(3\theta-{\pi\over 4} )+\sqrt{9-8\sin 2\theta-\sin 6\theta}\right)\right), \quad - {3 \pi\over 8} <\theta<{\pi \over 8}.
\label{et4}
\end{align}
The lower (upper) portion of the thimble is covered by $0<\theta<\pi/8$ and the upper (lower) portion by $-3 \pi/8 <\theta<0$ respectively (see Fig.~\ref{fig:comparison}). At $\theta=0$, 

For the higher modes $x_m$, again in the LFA, we obtain exactly the same formulae as displayed above, with an additional overall factor of $m$. 

\end{document}